\documentclass[11pt]{article}

\usepackage[english]{babel}

\usepackage[utf8]{inputenc}
\usepackage[T1]{fontenc}
\usepackage{microtype}
\usepackage{moresize}

\usepackage{csquotes,textcase,xspace}

\usepackage{geometry}
\usepackage{setspace}
\usepackage{graphicx}
\usepackage{xcolor}
\usepackage[font=small,labelfont=bf,labelsep=space]{caption}
\usepackage{subfigure}
\addto\captionsenglish{}
\addto\captionsenglish{}
\definecolor{JoliBleu}{rgb}{0,0.55,0.55}
\definecolor{JoliVert}{rgb}{0.15,0.6,0}
\definecolor{JoliRouge}{rgb}{0.86,0.08,0}
\definecolor{JoliJaune}{rgb}{1,0.75,0}
\definecolor{JoliGris}{rgb}{0.52,0.52,0.51}
\definecolor{myblue}{RGB}{26, 77, 116}
\definecolor{myorange}{RGB}{181, 116, 30}
\definecolor{mydarkorange}{RGB}{166, 88, 0}
\definecolor{mygreen}{RGB}{21, 124, 80}
\definecolor{myblack}{RGB}{43, 65, 82}
\definecolor{myred}{rgb}{0.5, 0.0, 0.13}

\usepackage{titlesec}
\titleformat{\section}[block]{\Large\boldmath\bfseries}{\thesection}{1em}{}
\titleformat{\subsection}[block]{\large\boldmath\bfseries}{\thesubsection}{0.5em}{}
\usepackage{appendix}

\makeatletter

\@addtoreset{equation}{section}
\makeatother

\usepackage[bottom]{footmisc}
\usepackage{fancyhdr}

\usepackage{titletoc}
\usepackage[nosort]{cite}
\newcommand{\doi}[2]{\href{http://dx.doi.org/#2}{#1}}

\usepackage{hyperref}
\hypersetup{colorlinks=true,
        pdfstartview=FitV,
        linkcolor= mydarkorange,
        citecolor= mydarkorange,
        urlcolor= JoliGris!60!black,
        hypertexnames=false,
        linktoc=page}

\usepackage{tikz}
\usetikzlibrary{calc}

\usepackage{amsmath,amssymb,amsfonts,dsfont}
\usepackage{mathrsfs}
\usepackage{physics}
\usepackage{ytableau}
\ytableausetup{boxsize=1.1em,centertableaux}
\usepackage{stmaryrd}
\usepackage{nicefrac}
\allowdisplaybreaks[1]
\usepackage{cases}

\usepackage{multirow}
\usepackage{booktabs}
\usepackage{pdflscape}
\usepackage{array}

\usepackage[shortlabels]{enumitem}

\newcommand{\A}{\ensuremath{\mathcal{A}}\xspace}
\newcommand{\F}{\ensuremath{\mathcal{F}}\xspace}
\renewcommand{\H}{\ensuremath{\mathcal{H}}\xspace}
\newcommand{\M}{\ensuremath{\mathcal{M}}\xspace}
\renewcommand{\P}{\ensuremath{\mathcal{P}}\xspace}
\newcommand{\J}{\ensuremath{\mathcal{J}}\xspace}
\renewcommand{\d}{\ensuremath{\mathrm{d}}\xspace}
\renewcommand{\H}{\ensuremath{\mathcal{H}}\xspace}

\renewcommand{\O}{\ensuremath{\mathrm{O}}\xspace}
\newcommand{\SL}{\ensuremath{\mathrm{SL}}\xspace}
\newcommand{\Odd}{\ensuremath{\mathrm{O}(d,d)}\xspace}

\renewcommand{\Tr}[1]{\ensuremath{\mathrm{Tr}\left(#1\right)}\xspace}

\newcommand{\veccomp}{\mbox{\sf u}\xspace}

\newcommand{\be}{\begin{equation}}
\newcommand{\ee}{\end{equation}}

\usepackage[affil-it]{authblk}
\def\preprint{}
\def\emails{}

\makeatletter
\def\@maketitle{%
  \newpage
  \null\hfill{November 2022}\\
  \null\hfill\texttt{\preprint}
  \vskip 4em%
  \begin{center}%
  \let \footnote \thanks
    {\LARGE\bfseries \@title \par}%
    \vskip 2.5em%
    {\large
      \lineskip .5em%
      \begin{center}
        \begin{minipage}{0.95\textwidth}
            \begin{tabular}[t]{c}%
            \@author
            \end{tabular}
        \end{minipage}    
      \end{center}\par}%
    \vskip 1em
    \emails
  \end{center}%
  \par
  \vskip 1.5em}
\makeatother

\title{U-duality and  $\alpha'$ corrections in three dimensions}
\author[1]{Camille Eloy}
\author[2]{Olaf Hohm}
\author[3,4]{Henning Samtleben}
\affil[1]{Theoretische Natuurkunde, Vrije Universiteit Brussel, and the International Solvay Institutes, Pleinlaan 2, B-1050 Brussels, Belgium}
\affil[2]{Institute for Physics, Humboldt University Berlin, Zum Gro{\ss}en Windkanal 6, D-12489 Berlin, Germany}
\affil[3]{ENSL, CNRS, Laboratoire de physique, F-69342 Lyon, France}
\affil[4]{Institut Universitaire de France (IUF)}
\def\preprint{HU-EP-22/37}
\def\emails{camille.eloy@vub.be, ohohm@physik.hu-berlin.de, henning.samtleben@ens-lyon.fr}

\begin{document}

\maketitle
\thispagestyle{empty}

\begin{abstract}

We consider the target space theory of bosonic and heterotic string theory to first order in $\alpha'$ 
compactified to three dimensions, using a formulation that is manifestly T-duality invariant under $\O(d,d,\mathbb{R})$ 
with $d=23$ and $d=7$, respectively. 
While the two-derivative supergravity exhibits a symmetry enhancement to the U-duality group $\O(d+1,d+1)$, 
the continuous group is known to be broken to $\O(d,d,\mathbb{R})$ by the first $\alpha'$ correction. 
We revisit this observation  by computing the full effective actions in three dimensions to first order in $\alpha'$ 
by dualizing the vector gauge fields. We give a formally $\O(d+1,d+1)$ invariant formulation by invoking 
a vector compensator, and we observe  a chiral pattern that allows one to reconstruct the bosonic action from the heterotic action. 
Furthermore, we obtain a particular massive deformation by integrating out the external $B$ field.
This induces a novel Chern-Simons term based on composite connections 
that, remarkably,  is $\O(d+1,d+1)$ invariant to leading order in the deformation parameter.

\end{abstract}

\newpage
\setcounter{page}{1} 

\tableofcontents

\section{Introduction}

Our goal in this paper is to explore some features arising in the interplay of higher-derivative $\alpha'$ corrections of the effective  actions 
of string and M-theory and the duality properties that these theories are expected to exhibit for certain backgrounds. 
Arguably the simplest duality property of string theory is T-duality, which states that theories  compactified on 
toroidal backgrounds T$^d$ related by $\O(d,d,\mathbb{Z})$ transformations are physically equivalent, 
even though these backgrounds may be radically different as ordinary geometries. 
This means that conventional Einstein-Hilbert gravity looks quite different on these backgrounds, yet `stringy gravity' supposedly cannot tell the difference. 
Even larger duality groups arise for particular theories and backgrounds. 
 In this paper we consider compactifications to three spacetime dimensions, which are particularly interesting 
 since in the supergravity limit the corresponding 
 T-duality group $\O(7,7)$ is enhanced to  larger groups, 
 and discrete subgroups of these  so-called U-duality groups are 
 conjectured to be dualities of the full string/M-theory \cite{Sen:1994wr,Hull:1994ys}.  
 Concretely, $\O(7,7)$ is generally 
 enhanced to $\O(8,8)$, which for type II string theory or M-theory 
 is then further enhanced to E$_{8,8}$ (a non-compact form of the largest finite-dimensional exceptional Lie group).   
The effect of $\alpha'$ corrections on the 
E$_{8,8}$ enhancement is difficult to study, since in type II string theory the corrections start at $\alpha'^{\,3}$ with  eight derivatives, 
but luckily in three dimensions there is the smaller U-duality group $\O(8,8)$ that can be discussed in bosonic and heterotic string theory whose 
$\alpha'$ corrections start with four derivatives. (For heterotic string theory the U-duality group is really $\O(8,24)$ but we truncate the vector fields, 
which reduces the group to $\O(8,8)$.)

At the level of the low-energy effective target space actions the T-duality property manifests itself 
in dimensional reduction (Kaluza-Klein compactification  on T$^d$ and subsequent truncation to massless modes), 
by exhibiting  a global symmetry under the continuous group $\O(d,d,\mathbb{R})$.\footnote{The enhancement to the continuous group can be understood as follows: the T-duality group $\O(d,d,\mathbb{Z})$ is discrete 
for purely geometrical reasons, because the symmetry transformations of fields need to be compatible with the periodicity conditions of 
the torus, but in dimensional reduction all memory of the torus has disappeared, explaining the enhancement to the continuous 
group. In contrast, the discrete T-duality group $\O(d,d,\mathbb{Z})$ is \textit{not} visible in the low-energy effective actions compactified on tori without truncation. This 
requires a genuine double field theory \cite{Hull:2009mi,Arvanitakis:2021ecw}.}  
The effective target space theories also receive an infinite number of higher-derivative corrections governed by the 
inverse string tension $\alpha'$, and it is known that the $\alpha'$ corrections 
preserve the continuous $\O(d,d,\mathbb{R})$ \cite{Sen:1991zi,Hohm:2014sxa} (see ref.~\cite{Hohm:2013bwa} for a review). 
Generalizing  previous work on cosmological reductions  to one dimension (cosmic time) \cite{Meissner:1996sa,Hohm:2015doa,Hohm:2019jgu}, 
we have recently determined the $\O(d,d,\mathbb{R})$ invariant effective actions to first order in $\alpha'$ for general reductions 
along $d$ dimensions \cite{Eloy:2019hnl,Eloy:2020dko}. 
We apply this effective action to compactifications to three dimensions, with the goal to explore the fate of the duality enhancement to $\O(d+1,d+1)$.
(Here $d=7$ for heterotic string theory and $d=23$ for bosonic string theory, but for our discussion $d$ is really a free parameter, 
and so we sometimes only speak of $\O(8,8)$ for the sake of vividness.)

In contrast to T-duality, which is a feature of classical string theory and preserved by all $\alpha'$ corrections, 
the U-dualities capture features of the quantum theory. Therefore, one should  perhaps not expect supergravity to exhibit 
U-duality symmetries beyond zeroth order in $\alpha'$ without also including quantum corrections. 
A simple argument based on a scaling symmetry of the two-derivative theory in fact shows that the \textit{continuous} 
U-duality group is not preserved to higher order in $\alpha'$ \cite{Lambert:2006he,Bao:2007er}. Indeed, the Einstein-Hilbert term in string frame,
\begin{equation}
  I_{\rm EH} = \int \d^{D}x\,\sqrt{g}\,e^{-\phi} R\,,
\end{equation}
has a global $\mathbb{R}\simeq\O(1,1)$ scaling symmetry with constant parameter $\lambda$, which acts as 
  $g_{\mu\nu} \rightarrow e^{\lambda} g_{\mu\nu}$ and 
  $\phi \rightarrow \phi + \frac{D-2}{2}\,\lambda$.   
This $\O(1,1)$  becomes  part of the U-duality group, but a typical higher-derivative coupling  of the form
\begin{equation}
  \alpha' \int \d^{D}x\,\sqrt{g}\,e^{-\phi} R_{\mu\nu\rho\sigma}R^{\mu\nu\rho\sigma}\,,
\end{equation}
is not invariant, but rather scales with $e^{-\lambda}$, hence breaking the symmetry. Intriguingly, however, the complete 
order $\alpha'$ action, including all matter couplings, scales homogeneously. There is hence a formal scaling invariance if one declares 
$\alpha'$ to scale with  $e^{+\lambda}$. Below we will employ a similar scheme to establish a formal $\O(8,8)$ invariance.

The above  scaling argument avoids 
the need to actually compute the effective action in, say, three dimensions, but by itself it is not sufficient to show that the \textit{discrete} subgroup 
is not realized in supergravity, because for the discrete group  the scaling symmetry in fact trivializes, as we will discuss. 
We therefore revisit the problem and compute the complete order $\alpha'$ effective action in three dimensions, starting from the results 
of ref.~\cite{Eloy:2019hnl,Eloy:2020dko} and using, perturbatively in $\alpha'$,  on-shell transformations to dualize  the vector gauge fields 
into scalar fields. In the two-derivative theory, this transformation shows  the enhancement from $\O(7,7)$ to $\O(8,8)$ (or E$_{8,8}$), 
because the new scalars 
organize into a larger coset matrix ${\cal M}_{\cal M\cal N}$, with $\O(8,8)$ indices ${\cal M}, {\cal N}\in\llbracket1, 16\rrbracket$. 
As expected, to first order in $\alpha'$ one finds that the continuous $\O(8,8)$ is no longer present, or at least no longer manifest. 
The manifest symmetry is $\O(7,7)$ times  $14$-dimensional translations that 
act as shifts on the scalars originating from dualization.

While $\O(8,8)$ is not a manifest symmetry we provide a formulation that, as alluded to above, exhibits  a formal  invariance under this group 
upon introducing a non-dynamical compensator. Specifically, for this we can choose  a constant vector $\veccomp^{\cal M}$ in the fundamental representation, 
in terms of  which the effective action takes  an $\O(8,8)$ invariant form. The true theory then  arises for a fixed vector 
$\veccomp^{\cal M}$ pointing in a particular direction. Even though it is in principle always possible to restore a broken symmetry in a formal manner 
by introducing an unphysical tensor compensator whose fictitious transformations absorb the failure of the actual theory to be invariant, 
this approach  may still be technically useful. For instance, it allows us to systematically determine the actual symmetry group 
as the invariance group of $\veccomp^{\cal M}$. Here  we indeed find the expected `geometric' subgroup, without any hidden 
enhancements. (Strictly speaking, this conclusion applies only to the continuous case, since in this formalism it is difficult to 
test for discrete symmetries without automatically testing for continuous symmetries.) 

More importantly, we find an intriguing `chiral pattern', in which only  one chiral projection of $\veccomp^{\cal M}$
is needed to write the action. To explain this, recall that given the $\O(8,8)$ coset matrix ${\cal M}$ and invariant metric $\eta$ 
one can define the projection operators 
\begin{equation}\label{projectorsIntro}
  P_{\cal MN} = \frac{1}{2}\left(\eta_{\cal MN}-\M_{\cal MN}\right)\;, \quad\quad \bar P_{\cal MN} = \frac{1}{2}\left(\eta_{\cal MN}+\M_{\cal MN}\right),
\end{equation}
onto two subspaces of opposite `chirality'. 
The four-derivative action turns out to be fully determined by an $\O(8,8)$ invariant function ${\cal F}$ of a 2-tensor and a vector argument, respectively, as follows
 \be\label{actionIntro}
  I_1 = \frac{1}{4}  \int\d^{3}x\,\sqrt{-g_{\rm E}}\,
  \Big\{ a\,{\cal F}[{\cal M},P\veccomp] + b\, \left({\cal F}[{\cal M},{P}\veccomp]\right)^{\star} \Big\}   \;. 
 \ee
Here `${}^{\star}$' denotes a $\mathbb{Z}_2$ action, which is implemented on the coset matrix as ${\cal M} \rightarrow {\cal Z}^{\rm t} {\cal M}{\cal Z}$, 
where ${\cal Z}$ obeys ${\cal Z}^2={\bf 1}$ but is \textit{not} an $\O(8,8)$ matrix. Under this $\mathbb{Z}_2$ the projectors (\ref{projectorsIntro}) 
are interchanged: 
 \be
  P \rightarrow {\cal Z}^{\rm t}\bar{P}{\cal Z} \;, \qquad \bar{P} \rightarrow {\cal Z}^{\rm t}{P}{\cal Z} \;. 
 \ee  
The parameters $a, b$ determine  the theory: the heterotic action is obtained for  $(a,b)=(-\alpha',0)$ and the bosonic action for $(a,b)=(-\alpha',-\alpha')$. 
The $\mathbb{Z}_2$ action, which exchanges $a$ and $b$, has a higher-dimensional analogue, sending the $B$ field $B\rightarrow -B$, which 
is a symmetry of the bosonic action  but not of the heterotic action. Since it is the same function ${\cal F}$ that determines the `$\mathbb{Z}_2$ dual' 
terms in the action, it follows that the bosonic action can be reconstructed from the heterotic action.

The above parameterization in terms of $(a,b)$, together with the $\mathbb{Z}_2$ 
action,  mimic the structure of  double field theory at order $\alpha'$ \cite{Hohm:2014xsa,Marques:2015vua}.  
The crucial `experimental' observation provided by our computation is that the compensator $\veccomp^{\cal M}$ in eq.~\eqref{actionIntro} 
appears only in a `chiral' or projected form or, alternatively, that there is the formal 
`gauge invariance' under $\veccomp \rightarrow \veccomp + \eta\bar{P}\Lambda$. 
(Note that this comes very close to an actual symmetry enhancement, with an $\O(8)$ acting only on indices with a `barred' projection, 
but this viewpoint is not quite consistent as the projectors are field dependent and hence not compatible with global symmetries.) 
We do not have an explanation  for this chiral pattern, but it is natural to conjecture 
that it holds to higher orders in $\alpha'$, hence providing indirect constraints on the allowed  higher-derivative couplings.

As the second main result of our paper we consider a particular massive deformation, as a window into more general gauged supergravities 
in presence of $\alpha'$ corrections. The latter would be important in order to study, for instance, the fate of Kaluza-Klein truncations on spheres
in presence of higher derivatives.  The massive deformation we consider is obtained by integrating out the external $B$ field.  In three dimensions 
its field strength is on-shell a constant that is usually set to zero in dimensional reduction. Keeping instead this constant $m$ while  
integrating out the $B$ field leads to a massive deformation, which for the two-derivative theory includes a potential term for the dilaton 
and a Chern-Simons term for the Kaluza-Klein vectors $A_{\mu}{}^{M}$~\cite{Kaloper:1993fg}. Including then the order $\alpha'$ corrections 
 one obtains additional couplings, which include a Chern-Simons term based on composite connections, 
with the latter originating from the scalar-dependent Green-Schwarz deformation uncovered in ref.~\cite{Eloy:2019hnl,Eloy:2020dko} (and given a worldsheet interpretation in ref.~\cite{Bonezzi:2020ryb}). 
Specifically, introducing an $\O(8,8)$ frame field ${\cal V}$, the compact part of its Maurer-Cartan form ${\cal V}^{-1}\d{\cal V} = {\cal P}+{\cal Q}$
defines \textit{composite} $\O(8)\times \O(8)$ connections ${\cal Q}$. The topological or Chern-Simons terms of the massive deformation 
at order $\alpha'$  and order $m$ then read 
 \be
  I_{(1) {\rm top}} = (a+b)\,m\int {\rm tr}\left({\cal Q}\wedge \d{\cal Q}+\frac{2}{3} {\cal Q} \wedge {\cal Q} \wedge {\cal Q} \right) 
  -\frac{a-b}{4}\, m \int {\rm tr} \left( \omega\wedge  \d\omega + \frac{2}{3} \omega \wedge \omega \wedge \omega\right) +{\cal O}(m^2) \,, 
 \ee
where $\omega$ denotes the Levi-Civita spin connection (so that for $a-b\neq 0$ this action includes topologically massive gravity 
as a subsector~\cite{Deser:1982vy}).  Unexpectedly, the Chern-Simons terms are hence  $\O(8,8)$ invariant to leading order in $m$, although the full theory is not. 
This is remarkable, for  there are now two parameters expected to break U-duality,  $\alpha'$ and $m$, 
yet for the leading Chern-Simons terms $\O(8,8)$ is restored. Again, we do not know what the physical significance 
of this observation is, and it remains to explore more general  gaugings. 

The remainder of this paper is organized as follows. In sec.~\ref{sec:I0} we give a short review of the duality enhancement in three dimensions 
for the two-derivative theory, with a particular focus on the scaling symmetries before and after dimensional reduction, 
since these feature prominently in the subsequent discussion of $\alpha'$  corrections. 
In sec.~\ref{sec:I1} we compute the effective action in three dimensions to first order in $\alpha'$ by perturbatively 
dualizing the vector gauge fields into scalars, and we exhibit the chiral pattern explained above. 
We then turn in sec.~\ref{sec:Bfield} to a massive deformation, that is obtained by integrating out the external $B$ field,  
with a focus on the resulting Chern-Simons terms for composite connections that exhibit an enhancement 
to the full U-duality group to first order in the mass parameter.  
We conclude with a short outlook in sec.~\ref{sec:ccl}, while  various identities and intermediate results are collected in appendices.

\section{Duality Enhancement in Three Dimensions} \label{sec:I0}

In this section we discuss some general aspects of the bosonic and heterotic string effective actions dimensionally reduced to three spacetime dimensions. 
We review the 
scaling symmetries in higher-dimensions (prior to any dimensional reduction) and in three dimensions, as a preparation for the discussion of 
duality enhancement from \Odd to $\O(d+1,d+1)$ that is expected to be a feature of string theory in three dimensions. 
We close the section with a discussion of field redefinitions, which are needed once higher-derivative corrections are included.

  \subsection{Scaling symmetries}  \label{sec:higherdimscaling}
  The bosonic parts of the bosonic and heterotic string effective actions coincide at the two-derivative order upon truncating the Yang-Mills gauge fields 
  of the heterotic theory. They describe the dynamics of a metric $\hat{g}_{\hat\mu\hat\nu}$, a two-form $\hat{B}_{\hat\mu\hat\nu}$ and a dilaton $\hat\phi$ in $D=26$ and $D=10$ dimensions, respectively. In the string frame, the two-derivative action is given by
  \begin{equation} \label{eq:stringD}
    I_{0}^{(D)}=\int\d^{D}X\,\sqrt{-\hat{g}}\,e^{-\hat{\phi}}\left(\hat{R}+\partial_{\hat{\mu}}\hat{\phi}\,\partial^{\hat{\mu}}\hat{\phi}-\frac{1}{12}\,\hat{H}_{\hat{\mu}\hat{\nu}\hat{\rho}}\hat{H}^{\hat{\mu}\hat{\nu}\hat{\rho}}\right),
  \end{equation}
  with the Ricci scalar $\hat{R}$ and the field-strength $\hat{H}_{\hat\mu\hat\nu\hat\rho}=3\,\partial_{[\hat\mu}\hat{B}_{\hat\nu\hat\rho]}$. It features several global scaling symmetries, with group $\mathbb{R}^{+}\simeq \O(1,1)$ and constant parameter $\lambda$, that we list in the following.
  \begin{itemize}
    \item Constant dilaton shift:
    \begin{equation} \label{eq:phiD}
      \left({\rm Dilaton}^{D}\right):\qquad\hat{\phi} \rightarrow \hat{\phi}+\lambda, \qquad\hat{g}_{\hat{\mu}\hat{\nu}} \rightarrow e^{2\lambda/(D-2)}\,\hat{g}_{\hat{\mu}\hat{\nu}}, \qquad\hat{B}_{\hat{\mu}\hat{\nu}} \rightarrow e^{2\lambda/(D-2)}\,\hat{B}_{\hat{\mu}\hat{\nu}}.
    \end{equation}

    \item On-shell ``trombone'' symmetry:
    \begin{equation}  \label{eq:trombD}
      \left({\rm Trombone}^{D}\right):\qquad\hat{\phi} \rightarrow \hat{\phi}, \qquad\hat{g}_{\hat{\mu}\hat{\nu}} \rightarrow e^{2\lambda}\,\hat{g}_{\hat{\mu}\hat{\nu}}, \qquad\hat{B}_{\hat{\mu}\hat{\nu}} \rightarrow e^{2\lambda}\,\hat{B}_{\hat{\mu}\hat{\nu}}, 
    \end{equation}
    which leaves invariant the equations of motion but rescales uniformly the action.

    \item Scaling of the internal volume: as we are interested in dimensional reductions down to three dimensions, we consider a splitting of the $D$-dimensional coordinates $X^{\hat{\mu}}$ into $\{x^{\mu},y^{m}\}$, with $\mu\in\llbracket1,3\rrbracket$ and $m\in\llbracket1,D-3\rrbracket$ and decompose the fields as in ref.~\cite{Maharana:1992my}:
    \begin{subequations}{\label{eq:ansatzdimreduc}}
      \begin{align}
      \hat{g}_{\hat{\mu}\hat{\nu}}&= \begin{pmatrix}
                                            g_{\mu\nu}+A_{\mu}^{(1)\,p}G_{pq}\,A_{\nu}^{(1)\,q} & A_{\mu}^{(1)\,p}G_{pn} \\[.5ex]
                                            G_{mp}A_{\nu}^{(1)\,p} & G_{mn}
                                    \end{pmatrix}, \label{eq:ansatzg} \\
      \hat{B}_{\hat{\mu}\hat{\nu}}&= \begin{pmatrix}
                                      B_{\mu\nu}-A_{[\mu}^{(1)\,m}A_{\nu]\,m}^{(2)}+A_{\mu}^{(1)\,m}\,B_{mn}\,A_{\nu}^{(1)\,n} & A^{(2)}_{\mu\,n}-B_{np}\,A_{\mu}^{(1)\,p} \\
                                      -A^{(2)}_{\nu\,m}+B_{mp}\,A_{\nu}^{(1)\,p} & B_{mn}
                                    \end{pmatrix}, \label{eq:ansatzB}\\
      e^{\hat{\phi}}&=\sqrt{\det\left(G_{mn}\right)}\,e^{\Phi}.
      \end{align}
    \end{subequations}
    With this decomposition, and keeping the dependence on all coordinates, the action~\eqref{eq:stringD} features the additional scaling symmetry
    \begin{equation} \label{eq:torusscaling}
      \left({\rm Volume}^{D-3}\right):\qquad
        \begin{cases}
        y^{m} \rightarrow e^{-\lambda}\,y^{m}, \\
        \Phi \rightarrow \Phi+(3-D)\lambda,\\
        g_{\mu\nu} \rightarrow g_{\mu\nu}, \\
        B_{\mu\nu}\rightarrow B_{\mu\nu}, \\
        \end{cases}
        \qquad
        \begin{cases}
        G_{mn} \rightarrow e^{2\lambda}\,G_{mn}, \\
        B_{mn} \rightarrow e^{2\lambda}\,B_{mn},\\
        A_{\mu}^{(1)\,m} \rightarrow e^{-\lambda}\,A_{\mu}^{(1)\,m}, \\
        A_{\mu\,m}^{(2)} \rightarrow e^{\lambda}\,A_{\mu\,m}^{(2)}, \\
        \end{cases}
    \end{equation}
    corresponding to the ${\rm GL}(1)$ subgroup of the ${\rm GL}(D-3)$ action on the internal coordinates.
  \end{itemize}

  These three scaling symmetries are summarized, in the three-dimensional variables of eq.~\eqref{eq:ansatzdimreduc}, in tab.~\ref{tab:Ddimscaling}. These scaling symmetries  are at the origin of three-dimensional symmetries essential to the duality enhancement.

  \begin{table}
    \renewcommand{\arraystretch}{1.5}
    \centering
    \begin{tabular}{c|ccccccc}
    & $e^{\Phi}$ & $g_{{\rm E}\,\mu\nu}$ & $B_{\mu\nu}$ & $G_{mn}$ & $B_{mn}$ & $A_{\mu}^{(1)\,m}$ & $A_{\mu\,m}^{(2)}$ \\\hline
    $\left({\rm Dilaton}^{D}\right)$ & $1/(D-2)$ & $0$ & $2/(D-2)$ & $2/(D-2)$ & $2/(D-2)$ & $0$ & $2/(D-2)$ \\
    $\left({\rm Trombone}^{D}\right)$ & $3-D$ & $2D-4$ & $2$ & $2$ & $2$ & $0$ & $2$ \\
    $\left({\rm Volume}^{D-3}\right)$ & $3-D$ & $2D-6$ & $0$ & $2$ & $2$ & $-1$ & $1$
    \end{tabular}
    \caption{Scaling behaviour of the fields~\eqref{eq:ansatzdimreduc} under the transformations~\eqref{eq:phiD}, \eqref{eq:trombD} and~\eqref{eq:torusscaling}. We display the charges $q$ of each field $\varphi$, representing the transformation $\varphi \rightarrow e^{q\lambda}\,\varphi$. For convenience, we have indicated the shifted dilaton $\Phi$ and Einstein frame metric $g_{{\rm E}\,\mu\nu}=e^{-2\Phi}\,g_{\mu\nu}$ rather than the dilaton $\hat{\phi}$ and the string frame metric $g_{\mu\nu}$.}
    \label{tab:Ddimscaling}
  \end{table}

  \subsection*{Scaling symmetries in three dimensions}
  We now consider the three-dimensional theory which follows from toroidal compactification of the action~\eqref{eq:stringD} on ${\rm T}^{D-3}$
  and a subsequent truncation to the zero-modes. 
  Using the parametrization~\eqref{eq:ansatzdimreduc} this amounts to neglecting the dependence on the internal coordinates $y^{m}$. We furthermore move to the Einstein frame by rescaling the metric, $g_{\mu\nu}\rightarrow g_{{\rm E}\,\mu\nu} = e^{-2\Phi} g_{\mu\nu}$, which yields  the action~\cite{Maharana:1992my}
  \begin{equation} \label{eq:I0Gld}
    \begin{aligned}
    I_{0}=&\int\d^{3}x\,\sqrt{-g_{\rm E}}\,\bigg(R_{\rm E}-\partial_{\mu}\Phi\,\partial^{\mu}\Phi-\frac{e^{-4\Phi}}{12}\,H_{\mu\nu\rho}H^{\mu\nu\rho} +\frac{1}{4}\,\Tr{\partial_{\mu}G\partial^{\mu}G^{-1}} \\
    & +\frac{1}{4} \Tr{G^{-1}\partial_{\mu}BG^{-1}\partial^{\mu}B}-\frac{e^{-2\Phi}}{4}\,F_{\mu\nu}^{(1)\,m}G_{mn}F^{(1)\,\mu\nu\,n}-\frac{e^{-2\Phi}}{4} \,H_{\mu\nu m}G^{mn}{H^{\mu\nu}}_{n}\bigg)\, ,
    \end{aligned}
  \end{equation}
  with 
   \begin{equation}   
   \begin{split}
    H_{\mu\nu\rho}&=3\,\partial_{[\mu}B_{\nu\rho]}-(3/2)\left(A^{(1)\,m}_{[\mu}F_{\nu\rho]\,m}^{(2)}+F_{[\mu\nu}^{(1)\,m}A_{\rho]\,m}^{(2)}\right)\,, \\
    H_{\mu\nu m} &= F_{\mu\nu\,m}^{(2)} - B_{mn}F_{\mu\nu}^{(1)\,n} \,, 
   \end{split}
   \end{equation}   
   where $F_{\mu\nu}^{(1)\,m}=\partial_{\mu}A_{\nu}^{(1)\,m}-\partial_{\nu}A_{\mu}^{(1)\,m}$ is the abelian field strength for $A_{\mu}^{(1)\,m}$, 
   and similarly for  $A_{\mu\,m}^{(2)}$. 
    This action is invariant  under the following scaling symmetries: 
  \begin{itemize}
    \item Constant dilaton shift:
    \begin{equation}  \label{eq:phi3}
    \left({\rm Dilaton}^{3}\right):\qquad
      \begin{cases}
      \Phi \rightarrow \Phi+\lambda,\\
      g_{{\rm E}\,\mu\nu} \rightarrow g_{{\rm E}\,\mu\nu}, \\
      B_{\mu\nu}\rightarrow e^{2\lambda}\,B_{\mu\nu}, \\
      \end{cases}
      \quad
      \begin{cases}
      G_{mn} \rightarrow G_{mn}, \\
      B_{mn} \rightarrow B_{mn},\\
      A_{\mu}^{(1)\,m} \rightarrow e^{\lambda}\,A_{\mu}^{(1)\,m}, \\
      A_{\mu\,m}^{(2)} \rightarrow e^{\lambda}\,A_{\mu\,m}^{(2)}. \\
      \end{cases}
    \end{equation}
    \item On-shell ``trombone'' symmetry:
    \begin{equation}  \label{eq:tromb3}
      \left({\rm Trombone}^{3}\right):\qquad
      \begin{cases}
      \Phi \rightarrow \Phi,\\
      g_{{\rm E}\,\mu\nu} \rightarrow e^{2\lambda}\,g_{{\rm E}\,\mu\nu}, \\
      B_{\mu\nu}\rightarrow e^{2\lambda}\,B_{\mu\nu}, \\
      \end{cases}
      \quad
      \begin{cases}
      G_{mn} \rightarrow G_{mn}, \\
      B_{mn} \rightarrow B_{mn},\\
      A_{\mu}^{(1)\,m} \rightarrow e^{\lambda}\,A_{\mu}^{(1)\,m}, \\
      A_{\mu\,m}^{(2)} \rightarrow e^{\lambda}\,A_{\mu\,m}^{(2)}. \\
      \end{cases}
    \end{equation}
    \item Internal rescaling:
    \begin{equation} \label{eq:Tdualscaling}
    \left(\text{T-duality}\right):\qquad
      \begin{cases}
        \Phi \rightarrow \Phi,\\
        g_{{\rm E}\,\mu\nu} \rightarrow g_{{\rm E}\,\mu\nu}, \\
        B_{\mu\nu}\rightarrow B_{\mu\nu}, \\
        \end{cases}
        \qquad
        \begin{cases}
        G_{mn} \rightarrow e^{2\lambda}\,G_{mn}, \\
        B_{mn} \rightarrow e^{2\lambda}\,B_{mn},\\
        A_{\mu}^{(1)\,m} \rightarrow e^{-\lambda}\,A_{\mu}^{(1)\,m}, \\
        A_{\mu\,m}^{(2)} \rightarrow e^{\lambda}\,A_{\mu\,m}^{(2)}, \\
        \end{cases}
    \end{equation}
    which corresponds to the $\O(1,1)$ subgroup of the T-duality group $\O(D-3,D-3)$.
  \end{itemize}

  Tab.~\ref{tab:3dimscaling} summarizes these symmetries. They do not directly arise from the reduction of the higher-dimensional scaling symmetries of sec.~\ref{sec:higherdimscaling}. Rather, the scaling symmetries in three dimensions originate from the mixing of the higher-dimensional ones:
  \begin{equation}  \label{eq:scalingDto3}
    \begin{aligned}
      \left({\rm Dilaton}^{3}\right) &= \left({\rm Dilaton}^{D}\right) + \frac{D-3}{D-2}\,\left({\rm Trombone}^{D}\right)-\left({\rm Volume}^{D-3}\right), \\
      \left({\rm Trombone}^{3}\right) &= \left({\rm Trombone}^{D}\right)-\left({\rm Volume}^{D-3}\right), \\
      \left(\text{T-duality}\right) &= (D-3)\,\left({\rm Dilaton}^{D}\right) - \frac{D-3}{D-2}\,\left({\rm Trombone}^{D}\right)+\left({\rm Volume}^{D-3}\right).
    \end{aligned}
  \end{equation}

  \begin{table}
    \renewcommand{\arraystretch}{1.5}
    \centering
    \begin{tabular}{c|ccccccc}
    & $e^{\Phi}$ & $g_{{\rm E}\,\mu\nu}$ & $B_{\mu\nu}$ & $G_{mn}$ & $B_{mn}$ & $A_{\mu}^{(1)\,m}$ & $A_{\mu\,m}^{(2)}$ \\\hline
    $\left({\rm Dilaton}^{3}\right)$ & $1$ & $0$ & $2$ & $0$ & $0$ & $1$ & $1$ \\
    $\left({\rm Trombone}^{3}\right)$ & $0$ & $2$ & $2$ & $0$ & $0$ & $1$ & $1$ \\
    $\left(\text{T-duality}\right)$ & $0$ & $0$ & $0$ & $2$ & $2$ & $-1$ & $1$
    \end{tabular}
    \caption{Scaling behaviour of the fields~\eqref{eq:ansatzdimreduc} under the transformations~\eqref{eq:phi3}--\eqref{eq:Tdualscaling}. We display the charges $q$ of each field $\varphi$, representing the transformation $\varphi \rightarrow e^{q\lambda} \varphi$.}
    \label{tab:3dimscaling}
  \end{table}

  \subsection{\texorpdfstring{$\O(d+1,d+1)$}{O(d+1,d+1)} enhancement} \label{sec:dualityenhancement}

  \paragraph{\boldmath T duality and \Odd}
  As already mentioned, the scaling symmetry~\eqref{eq:Tdualscaling} is part of the bigger T-duality symmetry group $\O(D-3,D-3)=\Odd$ (with $d=23$ and $d=7$ in the bosonic and heterotic cases, respectively). The invariance under \Odd is best displayed upon packaging the $d^{2}$ scalar fields $G_{mn}$ and $B_{mn}$ into the \Odd matrix
  \begin{equation} \label{eq:HOdd}
    {\cal H}_{MN} = \begin{pmatrix}
    G_{mn}-B_{mp}G^{pq}B_{qn} & B_{mp}G^{pn}\\[.5ex]
                      -G^{mp}B_{pn} & G^{mn}                 
                      \end{pmatrix},
  \end{equation}
  parametrizing the coset space $\Odd/\left(\O(d)\times\O(d)\right)$, and regrouping the vector fields $A^{(1)\,m}_{\mu}$ and $A^{(2)}_{\mu\,m}$ into a single \Odd vector
  \begin{equation} \label{eq:AOdd}
   {\cal A}_{\mu}{}^{M} = \begin{pmatrix}
                          A_{\mu}^{(1)\,m} \\
                          A_{\mu\,m}^{(2)}
                          \end{pmatrix}.
  \end{equation}
  The action~\eqref{eq:I0Gld} then takes the form~\cite{Maharana:1992my}
  \begin{equation} \label{eq:I0}
  \begin{split}
    I_{0}=\int\d^{3}x\,\sqrt{-g_{\rm E}}\Big(&R_{\rm E}-\partial_{\mu}\Phi\,\partial^{\mu}\Phi+\frac{1}{8}\,\Tr{\partial_{\mu}\H\,\partial^{\mu}\H^{-1}}
    \\
    &-\frac{1}{12}e^{-4\Phi} \,H_{\mu\nu\rho}H^{\mu\nu\rho} 
    -\frac{1}{4}\,e^{-2\Phi}\,\F_{\mu\nu}{}^{M}\H_{MN}\F^{\mu\nu\,N}\Big)\,,
  \end{split}  
  \end{equation}
  with the field-strengths $\F_{\mu\nu}{}^{M}=2\,\partial_{[\mu}\A_{\nu]}{}^{M}$ and $H_{\mu\nu\rho}=3\,\partial_{[\mu}B_{\nu\rho]}-(3/2)\,\A_{[\mu}{}^{M}\F_{\nu\rho]\,M}$. We choose the metric signature to be $(-1,1,1)$. The \Odd indices are raised and lowered using the \Odd-invariant metric 
  \begin{equation}  \label{eq:etamatrix} 
    \eta^{MN} = \begin{pmatrix}
                      0 & {\delta^{m}}_{n} \\
                      {\delta_{m}}^{n} & 0
                    \end{pmatrix}.
  \end{equation}
  In particular, $\H^{MN} = \eta^{MP}\H_{PQ}\eta^{QN}$ is the inverse of the generalised metric $\H_{MN}$.

  The action~\eqref{eq:I0} is invariant under the \Odd transformation
  \begin{equation}  \label{eq:Oddtransfo}
    g_{{\rm E}\,\mu\nu} \rightarrow g_{{\rm E}\,\mu\nu}, \quad \Phi \rightarrow \Phi, \quad \H_{MN} \rightarrow L_{M}{}^{P}L_{N}{}^{Q}\H_{PQ}, \quad \A_{\mu}{}^{M} \rightarrow L^{M}{}_{N}\A_{\mu}{}^{N}, \quad B_{\mu\nu} \rightarrow B_{\mu\nu},
  \end{equation}
  with $L_{M}{}^{N}\in\Odd$, \textit{i.e.}~$L_{M}{}^{P}L_{N}{}^{Q}\eta_{PQ} = \eta_{MN}$. 
  In three dimensions the three-form field-strength $H_{\mu\nu\rho}$ is on-shell determined by a constant, 
  which we set to zero for now. In sec.~\ref{sec:Bfield} we will explore the massive deformations arising for  a non-vanishing 
  three-form.

  \paragraph{\boldmath From \Odd to $\O(d+1,d+1)$}
  The action~\eqref{eq:I0} hides a symmetry enhancement from \Odd to $\O(d+1,d+1)$, thanks to the duality between vector and scalar fields in three dimensions~\cite{Sen:1994wr}. Contrary to T-duality, this enhanced symmetry does not leave the dilaton invariant: it combines in particular the $\O(1,1)$ scaling symmetry~\eqref{eq:phi3} to the \Odd symmetry~\eqref{eq:Oddtransfo} as $\Odd\times\O(1,1)\subset\O(d+1,d+1)$. The enhancement is made manifest by dualising the 2-form field-strengths $\F_{\mu\nu}{}^{M}$ into gradients $\partial_{\mu}\xi_{M}$ of scalar fields through the introduction of a Lagrange multiplier term in the action:\footnote{Here and in the following $\varepsilon_{\mu\nu\rho}$ denotes the Levi-Civita symbol, and $\epsilon_{\mu\nu\rho}$ is the associated tensor.}
   \begin{equation}
    \widetilde{I}_{0} = I_{0} + \int \d^{3}x
    \, \frac{1}{2}\,\varepsilon^{\mu\nu\rho}\F_{\mu\nu}{}^{M}\partial_{\rho}\xi_{M}\,.
  \end{equation}
  The equations of motion of $\xi_{M}$ give the Bianchi identity for $\F_{\mu\nu}{}^{M}$. There is therefore no need for the vector fields ${\cal A}_{\mu}{}^{M}$, and we can consider $\F_{\mu\nu}{}^{M}$ as independent fields. Their equations of motion, given by
  \begin{equation}  \label{eq:dual2der}
    \F_{\mu\nu}{}^{M} = e^{2\Phi}\,\epsilon_{\mu\nu\rho}\,\partial^{\rho}\xi_{N}\H^{NM}, 
  \end{equation}
  are algebraic: we can eliminate the 2-forms $\F_{\mu\nu}{}^{M}$ from the action in favour of the scalars $\xi_{M}$. The action then reeds
  \begin{equation} \label{eq:I0dual}
    \widetilde{I}_{0}=\int\d^{3}x\,\sqrt{-g_{\rm E}}\left(R_{\rm E}-\partial_{\mu}\Phi\,\partial^{\mu}\Phi+\frac{1}{8}\,\Tr{\partial_{\mu}\H\,\partial^{\mu}\H^{-1}}-\frac{1}{2}\,e^{2\Phi}\,\partial_{\mu}\xi_{M}\H^{MN}\partial^{\mu}\xi_{N}\right).
  \end{equation}
  The $\O(1,1)$ and \Odd transformations~\eqref{eq:phi3} and \eqref{eq:Oddtransfo} of $\A_{\mu}{}^{M}$ imply, by use of eq.~\eqref{eq:dual2der}, 
  the following transformations of  $\xi_{M}$:
  \begin{equation}
    \O(1,1):\ \xi_{M} \rightarrow e^{-\lambda}\,\xi_{M},\qquad \Odd:\ \xi_{M} \rightarrow L_{M}{}^{N}\xi_{N}.
  \end{equation}

  The action $\widetilde{I}_{0}$ depends on the metric and on $1+d^{2}+2d=(d+1)^{2}$ scalar fields. The scalar fields  can be organized into the $\O(d+1,d+1)$ matrix
  \begin{equation}  \label{eq:genmetd+1}
    \M_{\cal MN} = \begin{pmatrix}
                   \H_{MN} +e^{2\Phi}\xi_{M}\xi_{N} & e^{2\Phi} \xi_{M} & -\H_{MP}\xi^{P}-\dfrac{1}{2}e^{2\Phi}\xi_{M}\,\xi_{P}\xi^{P} \\
                   e^{2\Phi}\xi_{N} & e^{2\Phi} & -\dfrac{1}{2}e^{2\Phi}\xi_{P}\xi^{P} \\
                   -\H_{NP}\xi^{P}-\dfrac{1}{2}e^{2\Phi}\xi_{N}\,\xi_{P}\xi^{P} & -\dfrac{1}{2}e^{2\Phi}\xi_{P}\xi^{P} & e^{-2\Phi} + \xi_{P}\H^{PQ}\xi_{Q} + \dfrac{1}{4}e^{2\Phi}\left(\xi_{P}\xi^{P}\right)^{2}
                   \end{pmatrix}\,. 
  \end{equation}
Here the    $\O(d+1,d+1)$ indices are  split as $\M = \{M,+,-\}$  with respect to $\Odd\times\O(1,1)$, while 
the $\O(d+1,d+1)$-invariant metric takes the form
  \begin{equation} \label{eq:etad+1}
    \eta_{\cal MN} =\begin{pmatrix}
                      \eta_{MN} & 0 & 0 \\
                      0 & 0 & 1 \\
                      0 & 1 & 0
                    \end{pmatrix}.
  \end{equation}
  Then, the action~\eqref{eq:I0dual} becomes
  \begin{equation}  \label{eq:I0d+1}
    \widetilde{I}_{0}=\int \d^{3}x\,\sqrt{-g_{\rm E}}\left(R_{\rm E}+\frac{1}{8}\,\Tr{\partial_{\mu}\M\partial^{\mu}\M^{-1}}\right).
  \end{equation}
  It is manifestly invariant under the $\O(d+1,d+1)$ transformation
  \begin{equation}  \label{eq:Od+1d+1transfo}
    g_{{\rm E}\,\mu\nu} \rightarrow g_{{\rm E}\,\mu\nu}, \quad \M_{\cal MN} \rightarrow L_{\cal M}{}^{\cal P}L_{\cal N}{}^{\cal Q}\M_{\cal PQ},
  \end{equation}
  with $L_{\cal M}{}^{\cal N} \in \O(d+1,d+1)$.

  \paragraph{\boldmath $\O(d+1,d+1)$ transformations}
  Let us have a closer look at the $\O(d+1,d+1)$ symmetry. The generators $T^{\cal MN}$ of $\O(d+1,d+1)$ can be decomposed into \Odd components as
  \begin{equation}  \label{eq:O88gen}
    T^{\cal MN}=\left\{T^{MN},T^{M+},T^{M-},T^{+-}\right\}.
  \end{equation}
  The $T^{MN}$ generate the \Odd transformation
  \begin{equation}  \label{eq:Oddtransfoxi}
    g_{{\rm E}\,\mu\nu} \rightarrow g_{{\rm E}\,\mu\nu}, \quad \Phi \rightarrow \Phi, \quad \H_{MN} \rightarrow L_{M}{}^{P}L_{N}{}^{Q}\H_{PQ}, \quad \xi_{M} \rightarrow L_{M}{}^{N}\xi_{N},
  \end{equation}
  while $T^{+-}$ generates the $\O(1,1)$ scaling symmetry
  \begin{equation}  \label{eq:O11scaling}
    g_{{\rm E}\,\mu\nu} \rightarrow g_{{\rm E}\,\mu\nu}, \quad \Phi \rightarrow \Phi+\lambda,\quad \H_{MN} \rightarrow \H_{MN},\quad \xi_{M} \rightarrow e^{-\lambda}\,\xi_{M}.
  \end{equation}
  The charged generators $T^{M+}$ generate the constant shifts
  \begin{equation}  \label{eq:xishift}
    \xi_{M} \rightarrow \xi_{M}+c_{M}
  \end{equation}
  and the components $T^{M-}$ lead to complicated non-linear transformations.

  \paragraph{Frame formalism}
  For later convenience, let us define the frame fields
  \begin{equation}
    {\cal V}_{\cal M}{}^{\cal A} = \begin{pmatrix}
                                    E_{M}{}^{A} & e^{\Phi}\xi_{M} & 0 \\
                                    0 & e^{\Phi} & 0 \\
                                    -\xi^{P}E_{P}{}^{A} & -\dfrac{1}{2}e^{\Phi}\xi_{P}\xi^{P} & e^{-\Phi}
                                    \end{pmatrix},
  \end{equation}
  so that $\M_{\cal M N}={\cal V}_{\cal M}{}^{\cal A}\delta_{\cal AB}{\cal V}_{\cal N}{}^{\cal B}$, with the $(2d+2)\times(2d+2)$ identity matrix $\delta_{\cal AB}$. $E_{M}{}^{A}$ is the frame field associated to the \Odd generalized metric, \textit{i.e.} $\H_{MN}=E_{M}{}^{A}\delta_{AB}E_{N}{}^{B}$, and we denote the inverse by $E_{A}{}^{M}$. As for 
 the `curved' $\O(d+1,d+1)$ indices $\M$, we split the flat indices as ${\cal A} = \{A,\hat +,\hat -\}$. These flat indices are raised and lowered using the flat version of the invariant tensor~\eqref{eq:etad+1}:
  \begin{equation}
    \eta_{\cal AB}= {\cal V}_{\cal A}{}^{\cal M}\eta_{\cal MN}{\cal V}_{\cal B}{}^{\cal N} =\begin{pmatrix}
                              \eta_{AB} & 0 & 0 \\
                              0 & 0 & 1 \\
                              0 & 1 & 0
                            \end{pmatrix},
  \end{equation}
  with ${\cal V}_{\cal A}{}^{\cal M}$ the inverse frame field and $\eta_{AB}=E_{A}{}^{M}\eta_{MN}E_{B}{}^{N}$. The frame fields can be used to define the Maurer-Cartan form $\left({\cal V}^{-1}\partial_{\mu}{\cal V}\right)_{\cal A}{}^{\cal B} = {\cal P}_{\mu\,{\cal A}}{}^{B} + {\cal Q}_{\mu\,{\cal A}}{}^{B}$, where
  \begin{align}
    \P_{\mu\,{\cal A}}{}^{\cal B} & = \begin{pmatrix}
                                    P_{\mu\,A}{}^{B} & \dfrac{1}{2}e^{\Phi}E_{A}{}^{M}\partial_{\mu}\xi_{M} & -\dfrac{1}{2}e^{\Phi}\partial_{\mu}\xi^{M}E_{M}{}^{C}\delta_{CA} \\
                                    \dfrac{1}{2}e^{\Phi}\partial_{\mu}\xi_{M}E_{C}{}^{M}\delta^{CB} & \partial_{\mu}\Phi & 0 \\
                                    -\dfrac{1}{2}e^{\Phi}\partial_{\mu}\xi^{M}E_{M}{}^{B} & 0 & -\partial_{\mu}\Phi 
                                  \end{pmatrix}, \label{eq:Pd+1} \\
    {\cal Q}_{\mu\,{\cal A}}{}^{\cal B} & = \begin{pmatrix}
                                    Q_{\mu\,A}{}^{B} & \dfrac{1}{2}e^{\Phi}E_{A}{}^{M}\partial_{\mu}\xi_{M} & \dfrac{1}{2}e^{\Phi}\partial_{\mu}\xi^{M}E_{M}{}^{C}\delta_{CA} \\
                                    -\dfrac{1}{2}e^{\Phi}\partial_{\mu}\xi_{M}E_{C}{}^{M}\delta^{CB} & 0 & 0 \\
                                    -\dfrac{1}{2}e^{\Phi}\partial_{\mu}\xi^{M}E_{M}{}^{B} & 0 & 0
                                  \end{pmatrix}, \label{eq:Qd+1}
  \end{align}
  such that $\left(\P_{\mu}\delta\right)_{\cal AB}$ and $\left({\cal Q}_{\mu}\delta\right)_{\cal AB}$ are symmetric and antisymmetric, respectively. 
  Sometimes it is also convenient to use the basis in which $\eta_{\cal AB}$ is diagonal and has the form
  \begin{equation} \label{eq:etad+12}
   \eta_{\cal AB} =\begin{pmatrix}
                      - \delta_{ab} & 0 \\
                      0 & \delta_{\bar a\bar b}
                    \end{pmatrix},
  \end{equation}
  with the index split ${\cal A}\longrightarrow \{a,\bar a\},\ a,\bar a\in \llbracket 0,d \rrbracket$. In this basis, the Maurer-Cartan components  ${\cal P}_{\mu}{}^{\cal AB}$ can be expressed in terms of an $\O(d+1)$ bivector ${\mathsf{P}}_{\mu}{}^{a\bar a}$:
  \begin{equation} \label{eq:Pab}
   {\cal P}_{\mu}{}^{\cal AB} =\begin{pmatrix}
                      0 & \mathsf{P}_{\mu}{}^{a\bar b} \\
                      \mathsf{P}_{\mu}^{{\rm t}}{}^{\bar a b} & 0
                    \end{pmatrix}.
  \end{equation}

  \subsection{Field redefinitions}
  \label{sec:fieldredef}

  The presence of higher-derivative corrections makes it possible to perform field redefinitions that are perturbative in $\alpha'$, and previous works showed that they are necessary  to exhibit duality symmetries~\cite{Meissner:1996sa}. We denote by $\widetilde{I}_{1}$ the part of the action of order $\alpha'$ (after dualisation of the vector fields), so that  $\widetilde{I}=\widetilde{I}_{0}+\widetilde{I}_{1}+{\cal O}\left(\alpha'^{2}\right)$ is the total action. In the same manner as in ref.~\cite{Hohm:2015doa,Hohm:2019jgu}, we consider field redefinitions of the form
  \begin{equation}  \label{eq:fieldvariation}
    \varphi \rightarrow \varphi + \alpha'\,\delta\varphi,
  \end{equation}
  where $\varphi$ denotes a generic field. Under such redefinitions, the variation of $\widetilde{I}_{0}$ is
  \begin{equation}
    \delta \widetilde{I}_{0} = \alpha'\, \int \d^{3}x\,\sqrt{-g_{\rm E}}\bigg[\delta\Phi\,E_{\Phi} + \delta g^{\mu\nu}_{\rm E} E_{g\,\mu\nu}+ \Tr{\delta \H^{-1}\,E_{\cal H}}+E_{\xi}{}^{M}\delta\xi_{M}\bigg],
  \end{equation}
  where\footnote{Note that, as $\H\eta\H=\eta$, $\H$ is a constrained field and that the derivation of $E_{\H}$ by variation of the action must be done carefully~\cite{Hohm:2019jgu}.}
  \begin{subnumcases}{\label{eq:eom}}
    E_{\Phi} = 2\,\Box\Phi - e^{2\Phi}\,\partial_{\mu}\xi_{M}{\cal H}^{MN}\partial^{\mu}\xi_{N}, \label{eq:eomphi} \\
    E_{g\,\mu\nu} = R_{{\rm E}\,\mu\nu}-\partial_{\mu}\Phi\partial_{\nu}\Phi + \frac{1}{8}\,\Tr{\partial_{\mu}\H\,\partial_{\nu}\H^{-1}}-\frac{1}{2}\,e^{2\Phi}\,\partial_{\mu}\xi_{M}{\cal H}^{MN}\partial_{\nu}\xi_{N} \nonumber  \\ 
    \qquad \quad - \frac{1}{2}g_{{\rm E}\, \mu\nu} \left(R_{\rm E}-\partial_{\rho}\Phi\,\partial^{\rho}\Phi+\frac{1}{8}\,\Tr{\partial_{\rho}\H\,\partial^{\rho}\H^{-1}}-\frac{1}{2}\,e^{2\Phi}\,\partial_{\rho}\xi_{M}{\cal H}^{MN}\partial^{\rho}\xi_{N}\right), \label{eq:eomg} \\
    E_{{\cal H}\,MN}  = -\frac{1}{4}\,\bigg[\Box\,\H_{MN}+\left(\H\partial_{\mu}\H^{-1}\,\partial^{\mu}\H\right)_{MN} + e^{2\Phi}\,\partial_{\mu}\xi_{M}\partial^{\mu}\xi_{N} \nonumber \\ 
    \qquad \qquad - e^{2\Phi}\,\H_{MP}\partial_{\mu}\xi^{P}\partial^{\mu}\xi^{Q} \H_{QN}\bigg], \label{eq:eomS} \\
    E_{\xi}{}^{M} = e^{2\Phi}\,\left(\Box\xi_{N}{\cal H}^{NM}+2\,\partial_{\mu}\Phi\partial^{\mu}\xi_{N}{\cal H}^{NM}+\partial_{\mu}\xi_{N}\partial^{\mu}{\cal H}^{NM}\right)\,, \label{eq:eomchi}
  \end{subnumcases}
and $\Box=\nabla_{\mu}\nabla^{\mu}$. As we will not consider orders in $\alpha'$ higher than one, there is no need to compute how the redefinition affects corrections $\widetilde{I}_{1}$ to $\widetilde{I}_{0}$ of order ${\cal O}(\alpha')$: this variation will generate ${\cal O}(\alpha'^{2})$ terms. The expressions of the shifts $\delta\varphi$ can then be chosen  to cancel given terms in $\widetilde{I}_{1}$. In the following, we will use these redefinitions to cancel terms that contain as factors the leading two-derivative contributions from the field equations, as was done in ref.~\cite{Eloy:2020dko}. These factors can be replaced as follows:
  \begin{equation}  \label{eq:fieldredef}
    \begin{aligned}
    \Box\Phi&\longrightarrow Q_{\Phi} = \frac{1}{2}\,e^{2\Phi}\,\partial_{\mu}\xi_{M}{\cal H}^{MN}\partial^{\mu}\xi_{N}, \\
    R_{{\rm E}\,\mu\nu}&\longrightarrow Q_{g\,\mu\nu} = \partial_{\mu}\Phi\partial_{\nu}\Phi - \frac{1}{8}\,\Tr{\partial_{\mu}\H\,\partial_{\nu}\H^{-1}} +\frac{1}{2}\,e^{2\Phi}\,\partial_{\mu}\xi_{M}{\cal H}^{MN}\partial_{\nu}\xi_{N}, \\
    \Box\,\H_{MN}&\longrightarrow Q_{\H\,MN} = -\left(\H\,\partial_{\mu}\H^{-1}\partial^{\mu}\H\right)_{MN} - e^{2\Phi}\,\partial_{\mu}\xi_{M}\partial^{\mu}\xi_{N} + e^{2\Phi}\,\H_{MP}\partial_{\mu}\xi^{P}\partial^{\mu}\xi^{Q}\H_{QN}, \\
    \Box\xi_{M}&\longrightarrow Q_{\xi\,M} =  -2\,\partial_{\mu}\Phi\partial^{\mu}\xi_{M}-\partial_{\mu}\xi_{N}\left(\partial^{\mu}{\cal H}^{-1}{\cal H}\right)^{N}{}_{M}.\\
    \end{aligned}
  \end{equation}
  These replacements and the associated field redefinitions are summed up in tab.~\ref{tab:fieldredef}.

  \begin{table}[t!]
    \renewcommand{\arraystretch}{1.75}
    \centering
    \begin{tabular}{c|c|c}
    Term in the action & Field redefinitions & Replacement \\ \hline\hline
    $\alpha'\,X\Box \Phi$ & $\delta\Phi = -\dfrac{1}{2}\,X$ & $\alpha'\,XQ_{\Phi}$ \\
    $\alpha'\,X^{\mu\nu}R_{{\rm E}\,\mu\nu}$& $\delta g_{\rm E}^{\mu\nu} = -X^{(\mu\nu)}+\,g_{\rm E}^{\mu\nu}\,X_{\rho}{}^{\rho}$& $\alpha'\,X^{\mu\nu}Q_{g\,\mu\nu}$ \\
    $\alpha'\,\Tr{X\Box\H}$ & $\delta\H^{MN} = 4\,X^{MN}$ & $\alpha'\,\Tr{XQ_{\H}}$ \\
    $\alpha'\,X^{M}\Box\xi_{M}$ & $\delta\xi_{M} = -e^{-2\Phi}{\cal H}_{MN}X^{N}$ & $\alpha'\,X^{M}Q_{\xi\,M}$
    \end{tabular}
    \caption{Replacement rules for the terms carrying the leading two-derivative contribution from the field equations descending from the two-derivative action~\eqref{eq:I0dual} and associated field redefinitions. The explicit replacement rules are given in eq.~\eqref{eq:fieldredef}.}
    \label{tab:fieldredef}
  \end{table}

\section{Four-derivative Action in Three Dimensions}  \label{sec:I1}

We now revisit the symmetry enhancement of sec.~\ref{sec:I0} in the presence of first order $\alpha'$ corrections. 
To this end, we start from the manifestly \Odd invariant four-derivative action of ref.~\cite{Eloy:2020dko} and perform the dualization
of the vector fields as in eq.~(\ref{eq:dual2der}) above. We then express the resulting action in terms of $\O(d+1,d+1)$ quantities
upon introduction of a non-dynamical compensator which reveals an interesting structure of the action. 
Higher order corrections obstruct the $\O(d+1,d+1)$ symmetry enhancement of the two-derivative action 
as is most straightforwardly seen by tracking the fate of the scaling symmetries discussed in sec.~\ref{sec:I0}. 
Typical four-derivative corrections to the action~\eqref{eq:stringD} are of the form~\cite{Metsaev:1987zx}
  \begin{equation}
    I_{1}\propto \alpha'\,\int\d^{D}X\,\sqrt{-\hat{g}}\,e^{-\hat{\phi}}\left(\hat{R}_{\hat{\mu}\hat{\nu}\hat{\rho}\hat{\sigma}}\hat{R}^{\hat{\mu}\hat{\nu}\hat{\rho}\hat{\sigma}}-\frac{1}{2}\,\hat{R}_{\hat{\mu}\hat{\nu}\hat{\rho}\hat{\sigma}}\hat{H}^{\hat{\mu}\hat{\nu}\hat{\lambda}}\hat{H}^{\hat{\rho}\hat{\sigma}}{}_{\hat{\lambda}}+\ldots\right).
  \end{equation}
Under the transformations~\eqref{eq:phiD}, \eqref{eq:trombD} and~\eqref{eq:torusscaling}, the action $I_1$ transform homogeneously
with the charges\footnote{In eq.~\eqref{eq:I1chargeD} and~\eqref{eq:I1scaling}, we give the charges of $I_{1}$ under the trombone symmetries using as convention that the charges of $I_{0}^{(D)}$ and $I_{0}$ under these transformations are $0$, \textit{i.e.} we drop the global factor under which the lowest order equations of motion are rescaled.}
\begin{equation} \label{eq:I1chargeD}
  \renewcommand{\arraystretch}{1.5}
  \begin{array}{c|ccc}
    & \left({\rm Dilaton}^{D}\right) & \left({\rm Trombone}^{D}\right) & \left({\rm Volume}^{D-3}\right) \\\hline
    I_{1} & -2/(D-2) & -2 & 0
  \end{array} .  
\end{equation}
Both the dilaton shift and the trombone symmetries are broken by $\alpha'$ corrections. With the translation~\eqref{eq:scalingDto3}, 
the charges of $I_{1}$ under the three-dimensional symmetries~\eqref{eq:phi3}--\eqref{eq:Tdualscaling} are
\begin{equation} \label{eq:I1scaling}
  \renewcommand{\arraystretch}{1.5}
  \begin{array}{c|ccc}
    & \left({\rm Dilaton}^{3}\right) & \left({\rm Trombone}^{3}\right) & \left(\text{T-duality}\right) \\\hline
    I_{1} & -2 & -2 & 0
  \end{array}.   
\end{equation}
In particular, the T-duality scaling $\O(1,1)\subset\O(d,d)$ is preserved in presence of first order $\alpha'$ corrections, in agreement with the arguments of ref.~\cite{Sen:1991zi}. The presence of $\O(d,d)$ was explicitly verified in ref.~\cite{Eloy:2019hnl,Eloy:2020dko} at first order in $\alpha'$. The symmetry under the dilaton shift~\eqref{eq:phi3} however is broken, and so is the symmetry enhancement from $\O(d,d)$ to $\O(d+1,d+1)$. 
In the following, we formulate the $\alpha'$ corrections to the three-dimensional action~\eqref{eq:I0d+1} in terms of  the $\O(d+1,d+1)$ objects 
defined in the previous section, by introduction of a non-dynamical compensator.

\subsection{Dualisation of the vector fields}

We will treat the cases of the bosonic and heterotic string effective actions at the same time, using the notations of ref.~\cite{Marques:2015vua}. Our starting
point is the manifestly $\O(d,d)$ invariant action of ref.~\cite{Eloy:2020dko} 
\begin{equation}  \label{eq:Ifull}
  \begin{aligned}
    I=&\int\d^{3}x\,\sqrt{-g}\,e^{-\Phi}\Bigg[R+\partial_{\mu}\Phi\,\partial^{\mu}\Phi+\frac{1}{8}\,\Tr{\partial_{\mu}\H\,\partial^{\mu}\H^{-1}}-\frac{1}{4}\,\F_{\mu\nu}{}^{M}\,\H_{M N}\,\F^{\mu\nu\,N} \\
    &-\frac{a+b}{8}\Big(R_{\mu\nu\rho\sigma}R^{\mu\nu\rho\sigma} +\frac{1}{16}\,\Tr{\partial_{\mu}\H\partial_{\nu}\H^{-1}\partial^{\mu}\H\partial^{\nu}\H^{-1}} - \frac{1}{32}\, \Tr{\partial_{\mu}\H\partial_{\nu}\H^{-1}} \Tr{\partial^{\mu}\H\partial^{\nu}\H^{-1}} \\
    &\qquad\qquad+\frac{1}{8}\, {\F_{\mu\nu}}^{M}\H_{MN}\F_{\rho\sigma}^{N}\F^{\mu\rho\,P}\H_{PQ}\F^{\nu\sigma\,Q} -\frac{1}{2}\,{\F_{\mu\nu}}^{M}\H_{MN}\F^{\mu\rho\,N}\F^{\nu\sigma\,P}\H_{PQ}\F_{\rho\sigma}{}^{Q} \\
    &\qquad\qquad+\frac{1}{8}\,{\F_{\mu\nu}}^{M}{\F_{\rho\sigma}}_{M}\F^{\mu\rho\,N}{\F^{\nu\sigma}}_{N}  -\frac{1}{2}\, R_{\mu\nu\rho\sigma} \F^{\mu\nu\,M}\H_{MN}\F^{\rho\sigma\,N} \\ 
    &\qquad\qquad-\frac{1}{2}\,{\F_{\mu\nu}}^{M}\left(\H\partial_{\rho}\H^{-1}\partial^{\nu}\H\right)_{MN} \F^{\mu\rho\,N} + \frac{1}{4}\,\F^{\mu\rho\,M}\H_{MN}\F^{\nu}{}_{\rho}{}^{N}\Tr{\partial_{\mu}\H\partial_{\nu}\H^{-1}}\Big)\\
    &+\frac{a-b}{4}\Big(-\frac{1}{16}\,\Tr{\partial_{\mu}\H\partial^{\mu}\H^{-1}\partial_{\nu}\H\partial^{\nu}\H^{-1}\H\eta} - \frac{1}{16}\,\F_{\mu\nu}{}^{M}\F_{\rho\sigma\,M}\,\F^{\mu\nu\,P}\H_{PQ}\F^{\rho\sigma\,Q}\\
    &\qquad\qquad+\frac{1}{4}\,R^{\mu\nu\rho\sigma}\F_{\mu\nu}{}^{M}\F_{\rho\sigma\,M} +\frac{1}{8}\,\F_{\mu\nu}{}^{M}\left(\partial_{\rho}\H\partial^{\rho}\H^{-1}\right)_{M}{}^{N}\F^{\mu\nu}{}_{N}\\[-4pt]
    &\qquad\qquad+\frac{1}{4}\,\F_{\mu\nu}{}^{M}\left(\partial^{\mu}\H\partial_{\rho}\H^{-1}\right)_{M}{}^{N}\F^{\nu\rho}{}_{N}\Big)+{\cal O}\left(\alpha'^{2}\right)\Bigg],
  \end{aligned}
\end{equation}
having eliminated the 2-form $B_{\mu\nu}$ by virtue of its three-dimensional field equations as in (\ref{eq:I0dual}) above.
The bosonic and heterotic actions correspond to $(a,b)=(-\alpha',-\alpha')$ and $(a,b)=(-\alpha',0)$, respectively~\cite{Marques:2015vua}. 

In order to express this action in terms of $\O(d+1,d+1)$ covariant objects, we first switch to the Einstein frame
\begin{equation}
g_{\mu\nu}\rightarrow g_{{\rm E}\,\mu\nu} =  e^{-2\Phi}g_{\mu\nu}
\;,
\end{equation}
and use the fact that in three dimensions the Riemann tensor can be expressed as 
\begin{equation}
    R_{\,\mu\nu\rho\sigma} = S_{\mu\rho}g_{\nu\sigma}+S_{\nu\sigma}g_{\mu\rho}-S_{\mu\sigma}g_{\nu\rho}-S_{\nu\rho}g_{\mu\sigma},
\end{equation}
in terms of the Schouten tensor $S_{\mu\nu} = R_{\mu\nu}-\dfrac{1}{4}\,R\,g_{\mu\nu}$.

Next, we dualise the vectorial degrees of freedom to scalar ones. 
To this end, we proceed as we did for the two-derivative action and introduce a Lagrange multiplier term to the action:
  \begin{equation}  \label{eq:actiondual}
    \widetilde{I} = I + \frac{1}{2}\, \int \d^{3}x\, \varepsilon^{\mu\nu\rho}\F_{\mu\nu}{}^{M}\partial_{\rho}\xi_{M}.
  \end{equation}
The equations of motion of $\xi_{M}$ still gives the Bianchi identities for $\F_{\mu\nu}{}^{M}$. But, considering $\F_{\mu\nu}{}^{M}$ as an independent field, its equations of motion $\delta \widetilde{I}/\delta \F=0$ now contain corrections of order $\alpha'$. These equations are algebraic in $\F$ and can be solved perturbatively in $\alpha'$. The solution takes the form
  \begin{equation}  \label{eq:dual4der}
    \F_{\mu\nu}{}^{M} = \F^{(0)}_{\mu\nu}{}^{M} + \alpha'\,\F^{(1)}_{\mu\nu}{}^{M},
  \end{equation}
  where $\F^{(0)}_{\mu\nu}{}^{M}$ is solution of $\delta \widetilde{I}_{0}/\delta \F=0$, as given in eq.~\eqref{eq:dual2der}. The exact expression of $\F^{(1)}_{\mu\nu}{}^{M}$ is not necessary for our purpose, as we now show. Eq.~\eqref{eq:dual4der} is algebraic and can be introduced in the action~\eqref{eq:actiondual}, which, schematically, takes the following form:
  \begin{equation}
    \widetilde{I}\left(\F^{(0)} + \alpha'\,\F^{(1)}\right) = \widetilde{I}_{0}\left(\F^{(0)}\right) + I_{1}\left(\F^{(0)}\right) + \alpha'\,\F^{(1)}\frac{\delta \widetilde{I}_{0}}{\delta\F}\left(\F^{(0)}\right)+{\cal O}\left(\alpha'^{2}\right),
  \end{equation}
  where $I_{1}$ is the four-derivative action. The dependence on $\F^{(1)}$ is thus proportional to the equation of motion of $\F$ at order $\alpha'^{0}$, evaluated at its solution $\F^{(0)}$. Then, at first order in $\alpha'$, the corrections to the duality relation~\eqref{eq:dual2der} cancel out in the action, and the lowest order relation can be used to dualise the vectors.

Applying this procedure, after some computation, the four-derivative part of the action (\ref{eq:Ifull}) turns into
  \begin{equation}  \label{eq:I1einstein}
    \begin{aligned}
      \widetilde{I}_{1} = &\int\d^{3}x\,\sqrt{-g_{\rm E}}\,e^{-2\Phi}\,\Bigg[\\
      &-\frac{a+b}{8}\Big(\frac{1}{16}\,\Tr{\partial_{\mu}\H\partial_{\nu}\H^{-1}\partial^{\mu}\H\partial^{\nu}\H^{-1}}-\frac{1}{32}\,\Tr{\partial_{\mu}\H\partial_{\nu}\H^{-1}}\Tr{\partial^{\mu}\H\partial^{\nu}\H^{-1}} \\
      &\qquad+4\,\partial_{\mu}\Phi\partial^{\mu}\Phi\,\partial_{\nu}\Phi\partial^{\nu}\Phi +\frac{1}{4}\,e^{4\Phi}\,\partial_{\mu}\xi^{M}\partial_{\nu}\xi_{M}\partial^{\mu}\xi^{N}\partial^{\nu}\xi_{N}\\
      &\qquad-\frac{1}{4}\,e^{4\Phi}\,\partial_{\mu}\xi_{M}\H^{MN}\partial_{\nu}\xi_{N}\partial^{\mu}\xi_{P}\H^{PQ}\partial^{\nu}\xi_{Q}-\frac{1}{2}\,e^{4\Phi}\,\partial_{\mu}\xi_{M}\H^{MN}\partial^{\mu}\xi_{N}\partial_{\nu}\xi_{P}\H^{PQ}\partial^{\nu}\xi_{Q} \\
      &\qquad+\frac{1}{2}\,e^{2\Phi}\,\partial_{\mu}\xi_{M}\left(\H^{-1}\partial_{\nu}\H\partial^{\nu}\H^{-1}\right)^{MN}\partial^{\mu}\xi_{N} -\frac{1}{2}\,e^{2\Phi}\,\partial_{\mu}\xi_{M}\left(\H^{-1}\partial^{\mu}\H\partial_{\nu}\H^{-1}\right)^{MN}\partial^{\nu}\xi_{N} \\ 
      &\qquad-\frac{1}{4}\,e^{2\Phi}\,\Tr{\partial_{\mu}\H\partial^{\mu}\H^{-1}}\,\partial_{\nu}\xi_{M}\H^{MN}\partial^{\nu}\xi_{N} +\frac{1}{4}\,e^{2\Phi}\,\Tr{\partial_{\mu}\H\partial_{\nu}\H^{-1}}\,\partial^{\mu}\xi_{M}\H^{MN}\partial^{\nu}\xi_{N}\\
      &\qquad-2\,e^{2\Phi}\,\partial_{\mu}\Phi\partial_{\nu}\Phi\,\partial^{\mu}\xi_{M}\H^{MN}\partial^{\nu}\xi_{N}+4\,R_{{\rm E}\,\mu\nu}R_{\rm E}{}^{\mu\nu}-R_{\rm E}^{2} +4\,\Box\Phi\Box\Phi +8\,\Box\Phi\,\partial_{\mu}\Phi\partial^{\mu}\Phi \\[2pt]
      &\qquad+4\,\nabla_{\mu}\nabla_{\nu}\Phi\nabla^{\mu}\nabla^{\nu}\Phi-8\,\nabla_{\mu}\nabla_{\nu}\Phi\partial^{\mu}\Phi\,\partial^{\nu}\Phi-8\,R_{{\rm E}\,\mu\nu}\nabla^{\mu}\nabla^{\nu}\Phi +8\,R_{{\rm E}\,\mu\nu}\,\partial^{\mu}\Phi \partial^{\nu}\Phi \\[2pt]
      &\qquad-4\,R_{\rm E}\,\partial_{\mu}\Phi\partial^{\mu}\Phi-2\,e^{2\Phi}R_{\rm E\,\mu\nu}\,\partial^{\mu}\xi_{M}\H^{MN}\partial^{\nu}\xi_{N} +e^{2\Phi}R_{\rm E}\,\partial_{\mu}\xi_{M}\H^{MN}\partial^{\mu}\xi_{N} \\[2pt]
      &\qquad-2\,e^{2\Phi}\Box\Phi\,\partial_{\mu}\xi_{M}\H^{MN}\partial^{\mu}\xi_{N} +2\,e^{2\Phi} \nabla_{\mu} \nabla_{\nu}\Phi\,\partial^{\mu}\xi_{M}\H^{MN}\partial^{\nu}\xi_{N}\Big)\\
      &+\frac{a-b}{4}\Big(-\frac{1}{16}\,\Tr{\partial_{\mu}\H\partial^{\mu}\H^{-1}\partial_{\nu}\H\partial^{\nu}\H^{-1}\H\eta} - \frac{1}{4}\,e^{4\Phi}\,\partial_{\mu}\xi_{M}\partial_{\nu}\xi^{M}\,\partial^{\mu}\xi_{P}\H^{PQ}\partial^{\nu}\xi_{Q}\\
      &\qquad-\frac{1}{2}\,e^{2\Phi}\,R_{\rm E}\,\partial_{\mu}\xi_{M}\partial^{\mu}\xi^{M} +e^{2\Phi}\,R_{\rm E\,\mu\nu}\,\partial^{\mu}\xi_{M}\partial^{\nu}\xi^{M}-e^{2\Phi}\,\nabla_{\mu}\nabla_{\nu}\Phi\,\partial^{\mu}\xi_{M}\partial^{\nu}\xi^{M}\\[-5pt]
      &\qquad+e^{2\Phi}\,\Box\Phi\,\partial_{\mu}\xi_{M}\partial^{\mu}\xi^{M}+e^{2\Phi}\,\partial_{\mu}\Phi\partial_{\nu}\Phi\,\partial^{\mu}\xi_{M}\partial^{\nu}\xi^{M}-\frac{1}{4}\,e^{2\Phi}\,\partial_{\mu}\xi_{M}\left(\partial_{\nu}\H^{-1}\partial^{\mu}\H\right)^{M}{}_{N}\partial^{\nu}\xi^{N}\Big)\Bigg].
 \end{aligned}
\end{equation}
  We now convert all second order derivatives in eq.~\eqref{eq:I1einstein} into products of first order derivatives, to allow comparison with the basis of $\O(d+1,d+1)$-invariant four-derivative terms of ref.~\cite{Eloy:2020dko}. To do so, we first use integrations by parts so that all the second order derivatives appear in the leading two-derivative contribution of the equations of motion~\eqref{eq:eom}.  Using furthermore the field redefinitions of sec.~\ref{sec:fieldredef}, we obtain the action
  \begin{align}
    \widetilde{I}_{1} & = \int\d^{3}x\,\sqrt{-g_{\rm E}}\,e^{-2\Phi}\,\Bigg[\nonumber\\
    &-\frac{a+b}{8}\Big(-\partial_{\mu}\Phi\partial^{\mu}\Phi\,\partial_{\nu}\Phi\partial^{\nu}\Phi+\frac{1}{16}\,\Tr{\partial_{\mu}\H\partial_{\nu}\H^{-1}\partial^{\mu}\H\partial^{\nu}\H^{-1}} \nonumber\\
    &\qquad+\frac{1}{32}\,\Tr{\partial_{\mu}\H\partial_{\nu}\H^{-1}}\Tr{\partial^{\mu}\H\partial^{\nu}\H^{-1}}-\frac{1}{64}\,\Tr{\partial_{\mu}\H\partial^{\mu}\H^{-1}}\Tr{\partial_{\nu}\H\partial^{\nu}\H^{-1}}\nonumber\\
    &\qquad+\frac{1}{4}\,e^{4\Phi}\,\partial_{\mu}\xi^{M}\partial_{\nu}\xi_{M}\partial^{\mu}\xi^{N}\partial^{\nu}\xi_{N}-\frac{1}{4}\,e^{4\Phi}\,\partial_{\mu}\xi_{M}\H^{MN}\partial_{\nu}\xi_{N}\partial^{\mu}\xi_{P}\H^{PQ}\partial^{\nu}\xi_{Q} \nonumber\\
    &\qquad+\frac{1}{4}\,e^{4\Phi}\,\partial_{\mu}\xi_{M}\H^{MN}\partial^{\mu}\xi_{N}\partial_{\nu}\xi_{P}\H^{PQ}\partial^{\nu}\xi_{Q}+\frac{1}{2}\,e^{2\Phi}\,\partial_{\mu}\xi_{M}\left(\H^{-1}\partial_{\nu}\H\partial^{\nu}\H^{-1}\right)^{MN}\partial^{\mu}\xi_{N}  \nonumber\\ 
    &\qquad-\frac{1}{2}\,e^{2\Phi}\,\partial_{\mu}\xi_{M}\left(\H^{-1}\partial^{\mu}\H\partial_{\nu}\H^{-1}\right)^{MN}\partial^{\nu}\xi_{N}-2\,e^{2\Phi}\,\partial_{\mu}\Phi\partial_{\nu}\Phi\,\partial^{\mu}\xi_{M}\H^{MN}\partial^{\nu}\xi_{N} \nonumber\\
    &\qquad+\frac{1}{2}\,\partial_{\mu}\Phi\partial_{\nu}\Phi\,\Tr{\partial^{\mu}\H\partial^{\nu}\H^{-1}}-\frac{1}{4}\,\partial_{\mu}\Phi\partial^{\mu}\Phi\,\Tr{\partial_{\nu}\H\partial^{\nu}\H^{-1}}  +e^{2\Phi}\,\partial_{\mu}\Phi \,\partial_{\nu}\xi_{M}\partial^{\mu}\H^{MN}\partial^{\nu}\xi_{N} \Big)\nonumber\\
    &+\frac{a-b}{4}\Big(-\frac{1}{16}\,\Tr{\partial_{\mu}\H\partial^{\mu}\H^{-1}\partial_{\nu}\H\partial^{\nu}\H^{-1}\H\eta} + \frac{1}{4}\,e^{4\Phi}\,\partial_{\mu}\xi_{M}\partial_{\nu}\xi^{M}\,\partial^{\mu}\xi_{P}\H^{PQ}\partial^{\nu}\xi_{Q}\nonumber\\
    &\qquad-\frac{1}{4}\,e^{2\Phi}\,\partial_{\mu}\xi_{M}\left(\partial_{\nu}\H^{-1}\partial^{\mu}\H\right)^{M}{}_{N}\partial^{\nu}\xi^{N}-\frac{1}{2}\,e^{2\Phi}\,\partial_{\mu}\Phi\partial^{\mu}\Phi\,\partial_{\nu}\xi_{M}\partial^{\nu}\xi^{M} -\frac{1}{8}\,\Tr{\partial_{\mu}\H\partial_{\nu}\H^{-1}}\,\partial^{\mu}\xi_{M}\partial^{\nu}\xi^{M}\nonumber\\[-5pt]
    &\qquad+\frac{1}{16}\,\Tr{\partial_{\mu}\H\partial^{\mu}\H^{-1}}\,\partial_{\nu}\xi_{M}\partial^{\nu}\xi^{M}+e^{2\Phi}\,\partial_{\mu}\Phi\,\partial^{\mu}\xi_{M}\left(\partial_{\nu}\H^{-1}\H\right)^{M}{}_{N}\partial^{\nu}\xi^{N}\Big)\Bigg]. \label{eq:I1firstderiv}
  \end{align}
  Explicitly, we have used the following order $\alpha'$ field redefinitions:
  \begin{equation}
    \begin{aligned}
      \delta g_{\rm E}^{\mu\nu} &= -\dfrac{a+b}{8\,\alpha'} \Big(-4\,e^{-2\Phi}\,R_{\rm E}^{\mu\nu}+2\,e^{-2\Phi}\,g_{\rm E}^{\mu\nu}\,R_{\rm E}+8\,e^{-2\Phi}\,\partial^{\mu}\Phi\partial^{\nu}\Phi-2\,e^{-2\Phi}\,g_{\rm E}^{\mu\nu}\,\partial_{\rho}\Phi\partial^{\rho}\Phi\\
      &\quad\qquad+\dfrac{1}{2}\,e^{-2\Phi}\,\Tr{\partial^{\mu}\H\partial^{\nu}\H^{-1}}-\dfrac{1}{4}\,e^{-2\Phi}\,g_{\rm E}^{\mu\nu}\,\Tr{\partial_{\rho}\H\partial^{\rho}\H^{-1}}-3\,g_{\rm E}^{\mu\nu}\,\partial_{\rho}\xi_{M}\H^{MN}\partial^{\rho}\xi_{N}\Big)\\
      &\quad-\dfrac{a-b}{4\,\alpha'}\,\partial^{\mu}\xi_{M}\partial^{\nu}\xi^{M},\\[5pt]
      \delta\Phi&=-\dfrac{a+b}{8\,\alpha'} \left(2\,e^{-2\Phi}\,R_{\rm E}-4\,e^{-2\Phi}\,\Box\Phi-\dfrac{3}{2}\,\partial_{\mu}\xi_{M}\H^{MN}\partial^{\mu}\xi_{N}\right)- \dfrac{a-b}{4\,\alpha'}\,\dfrac{1}{4}\,\partial_{\mu}\xi_{M}\partial^{\mu}\xi^{M},\\[5pt]
      \delta\xi_{M}&=-\dfrac{a+b}{4\,\alpha'}\,e^{-2\Phi}\,\partial_{\mu}\Phi\partial^{\mu}\xi_{M}-\dfrac{a-b}{4\,\alpha'}\,e^{-2\Phi}\,\H_{MN}\,\partial_{\mu}\Phi\partial^{\mu}\xi^{N},
    \end{aligned}
  \end{equation}
  in the notations of eq.~\eqref{eq:fieldvariation}.

\subsection{$\O(d+1,d+1)$-covariant formulation} \label{sec:formalcovariance}
We now aim to express the action (\ref{eq:I1firstderiv}) in terms of $\O(d+1,d+1)$-covariant objects. We define the $\O(d+1,d+1)$ currents $\J_{\mu} = \partial_{\mu}\M\M^{-1}$ for the matrix $\M$ from eq.~\eqref{eq:genmetd+1}. Explicitly, this current takes the form
\begin{equation}
\J_{\mu\,{\cal M}}{}^{{\cal N}}=
\begin{pmatrix}
 \J_{\mu\,M}{}^{N}
 &   \J_{\mu\,M}{}^{+} &  -e^{2\Phi}\partial_{\mu}\xi_{P}\H^{P}{}_{M}  \\
e^{2\Phi}\partial_{\mu}\xi_{P}\H^{PN}  &  2\partial_{\mu}\Phi-e^{2\Phi}\xi_{P}\H^{PQ}\partial_{\mu}\xi_{Q} & 0 \\
\J_{\mu\,-}{}^{N} & 0 &  -2\partial_{\mu}\Phi + e^{2 \Phi} \xi_{P}\H^{P Q}\partial_{\mu}\xi_{Q}
\end{pmatrix}
\;,
\end{equation}
with
\begin{equation}
    \begin{aligned}
      \J_{\mu\,M}{}^{N} &= \partial_{\mu}\H_{M P}\H^{P N} + e^{2\Phi} \left(\xi_{M}\partial_{\mu}\xi_{P} \H^{PN}-\H_{M P}\partial_{\mu}{\xi^{P}}\xi^{N}\right), \\
      \J_{\mu\,M}{}^{+} &= \partial_{\mu}\xi_{M}-\partial_{\mu}\H_{MP}\H^{PQ}\xi_{Q} + 2\,\partial_{\mu}\Phi\xi_{M}- e^{2 \Phi}\xi_{P}\H^{PQ}\partial_{\mu}\xi_{Q}\xi_{M} +\frac{1}{2}\,e^{2 \Phi} \xi_{P}\xi^{P}\partial_{\mu}\xi_{Q}\H^{Q}{}_{M} \ ,\\ 
      \J_{\mu\,-}{}^{N} &=-\partial_{\mu}\xi^{N}-\xi^{P}\partial_{\mu}\H_{PQ}\H^{QN} - 2\partial_{\mu}\Phi\xi^{N}+ e^{2\Phi}\xi_{P}\H^{PQ}\partial_{\mu}\xi_{Q}\xi^{N}-\frac{1}{2}e^{2\Phi}\xi_{P}\xi^{P}\partial_{\mu}\xi_{Q}\H^{Q N}\ . \\
    \end{aligned}
\end{equation}
In terms of this object and using the \Odd decomposition of app.~\ref{app:Od1d1toOdd}, the four-derivative action (\ref{eq:I1firstderiv}) can be cast into the rather compact form
\begin{equation}  \label{eq:I1d+1currentsproj}
  \begin{aligned}
  \widetilde{I}_{1} &= \int\d^{3}x\,\sqrt{-g_{\rm E}}\,e^{-2\Phi}\,\Bigg\{\\ 
  &+\frac{a}{4}\Bigg[-\frac{1}{32}\Tr{{\cal J}_{\mu}{\cal J}_{\nu}{\cal J}^{\mu}{\cal J}^{\nu}}-\frac{1}{16}\,\Tr{\J_{\mu}\J^{\mu}\J_{\nu}\J^{\nu}\M\eta}\\
  & \quad\ \,\quad-\frac{1}{64}\,\Tr{{\cal J}_{\mu}{\cal J}_{\nu}}\Tr{{\cal J}^{\mu}{\cal J}^{\nu}}+\frac{1}{128}\,\Tr{{\cal J}_{\mu}{\cal J}^{\mu}}\Tr{{\cal J}_{\nu}{\cal J}^{\nu}}\\
  &\quad\ \,\quad +e^{-2\Phi}\bigg(-\frac{1}{2}\,\left(\veccomp P\eta{\cal J}_{\mu}{\cal J}_{\nu}{\cal J}^{\mu}{\cal J}^{\nu}P\veccomp\right)+\frac{1}{2}\,\left(\veccomp P\eta{\cal J}_{\mu}{\cal J}^{\mu}{\cal J}_{\nu}{\cal J}^{\nu}P\veccomp\right)-\frac{1}{2}\,\left(\veccomp P\eta{\cal J}_{\mu}{\cal J}_{\nu}{\cal J}^{\nu}{\cal J}^{\mu}P\veccomp\right)\\
  &\qquad\ \,\qquad\qquad-\frac{1}{4}\,\Tr{{\cal J}_{\mu}{\cal J}_{\nu}}\left(\veccomp P\eta{\cal J}^{\mu}{\cal J}^{\nu}P\veccomp\right)+\frac{1}{8}\,\Tr{{\cal J}_{\mu}{\cal J}^{\mu}}\left(\veccomp P\eta{\cal J}_{\nu}{\cal J}^{\nu}P\veccomp\right)\bigg)\\
  &\quad\ \,\quad +e^{-4\Phi}\bigg(-2\,\left(\veccomp P\eta{\cal J}_{\mu}{\cal J}_{\nu}P \veccomp\right)\left(\veccomp P\eta{\cal J}^{\mu}{\cal J}^{\nu}P\veccomp\right)+\left(\veccomp P\eta{\cal J}_{\mu}{\cal J}^{\mu}P\veccomp\right)\left(\veccomp P\eta{\cal J}_{\nu}{\cal J}^{\nu}P\veccomp\right)\bigg)\Bigg]\\
  &+\frac{b}{4}\Bigg[-\frac{1}{32}\Tr{{\cal J}_{\mu}{\cal J}_{\nu}{\cal J}^{\mu}{\cal J}^{\nu}}+\frac{1}{16}\,\Tr{\J_{\mu}\J^{\mu}\J_{\nu}\J^{\nu}\M\eta}\\
  & \quad\ \,\quad-\frac{1}{64}\,\Tr{{\cal J}_{\mu}{\cal J}_{\nu}}\Tr{{\cal J}^{\mu}{\cal J}^{\nu}}+\frac{1}{128}\,\Tr{{\cal J}_{\mu}{\cal J}^{\mu}}\Tr{{\cal J}_{\nu}{\cal J}^{\nu}}\\
  &\quad\ \,\quad +e^{-2\Phi}\bigg(\frac{1}{2}\,\left(\veccomp \bar P\eta{\cal J}_{\mu}{\cal J}_{\nu}{\cal J}^{\mu}{\cal J}^{\nu}\bar P\veccomp\right)-\frac{1}{2}\,\left(\veccomp \bar P\eta{\cal J}_{\mu}{\cal J}^{\mu}{\cal J}_{\nu}{\cal J}^{\nu}\bar P\veccomp\right)+\frac{1}{2}\,\left(\veccomp \bar P\eta{\cal J}_{\mu}{\cal J}_{\nu}{\cal J}^{\nu}{\cal J}^{\mu}\bar P\veccomp\right)\\
  &\qquad\ \,\qquad\qquad+\frac{1}{4}\,\Tr{{\cal J}_{\mu}{\cal J}_{\nu}}\left(\veccomp \bar P\eta{\cal J}^{\mu}{\cal J}^{\nu}\bar P\veccomp\right)-\frac{1}{8}\,\Tr{{\cal J}_{\mu}{\cal J}^{\mu}}\left(\veccomp \bar P\eta{\cal J}_{\nu}{\cal J}^{\nu}\bar P\veccomp\right)\bigg)\\
  &\quad\ \,\quad +e^{-4\Phi}\bigg(-2\,\left(\veccomp \bar P\eta{\cal J}_{\mu}{\cal J}_{\nu}\bar P \veccomp\right)\left(\veccomp \bar P\eta{\cal J}^{\mu}{\cal J}^{\nu}\bar P\veccomp\right)+\left(\veccomp \bar P\eta{\cal J}_{\mu}{\cal J}^{\mu}\bar P\veccomp\right)\left(\veccomp \bar P\eta{\cal J}_{\nu}{\cal J}^{\nu}\bar P\veccomp\right)\bigg)\Bigg]\Bigg\}.
 \end{aligned}
\end{equation}
Here, we have defined the projectors
\begin{equation}
  P_{\cal MN} = \frac{1}{2}\,\left(\eta_{\cal MN}-\M_{\cal MN}\right)\quad{\rm and}\quad \bar P_{\cal MN} = \frac{1}{2}\,\left(\eta_{\cal MN}+\M_{\cal MN}\right),
\end{equation}
and the $\O(d+1,d+1)$ compensator vector 
\begin{equation}
\veccomp^{\cal M}= \{0,1,0\}
\;,
\label{veccomp}
\end{equation}
which parametrizes the breaking of the symmetry group $\O(d+1,d+1)$ at the four-derivative order.
The action (\ref{eq:I1d+1currentsproj}) enjoys a formal $\O(d+1,d+1)$
invariance, which is broken by the explicit choice (\ref{veccomp}) for the vector $\veccomp^{\cal M}$.
This is manifest in the above form for all terms except the explicit dilaton prefactor, 
but it even holds for these factor thanks to the relations (\ref{e2PhiRel}) below. 

It is interesting to note that  the above action (\ref{eq:I1d+1currentsproj}) has the chiral structure 
 \be
  I_1 = \frac{1}{4}  \int\d^{3}x\,\sqrt{-g_{\rm E}}\,
  \Big\{ a\,{\cal F}[{\cal M},P\veccomp] + b\, \left({\cal F}[{\cal M},{P}\veccomp]\right)^*\Big\}   
  \label{chiral}
 \ee
with a fixed  function ${\cal F}$ that depends only on ${\cal M}$ and a projection of $\veccomp$. 
The `$^*$' in the second term indicates the $\mathbb{Z}_2$ action, under which ${\cal M}$ transforms as 
 \be
  {\cal M} \rightarrow {\cal Z}^{\rm t} {\cal M}{\cal Z}, \qquad {\cal Z} = \begin{pmatrix} Z && 0 \\ 0 && \sigma_3 \end{pmatrix}, 
  \qquad  Z = \begin{pmatrix} {\bf 1} && 0 \\ 0 && -{\bf 1} \end{pmatrix},
 \ee
where 
 \be
  {\cal Z}^{\rm t} \eta{\cal Z}=-\eta\;. 
 \ee
Consequently,  the projectors transform as 
 \be
  P \rightarrow {\cal Z}^{\rm t} \bar{P}{\cal Z} \;, \qquad \bar{P} \rightarrow {\cal Z}^{\rm t} {P}{\cal Z} 
  \;,
 \ee  
while furthermore
 \be\label{e2PhiRel}
  e^{2\Phi} = \veccomp^M {\cal M}_{MN}\, \veccomp^N = -2  \veccomp^M {P}_{MN}\, \veccomp^N=2  \veccomp^M \bar{P}_{MN}\, \veccomp^N
  \;.
 \ee
 In particular, the last equation shows that the dilaton $\Phi$ is invariant under the $\mathbb{Z}_2$ action.
The chiral form (\ref{chiral}) of the action implies that the four-derivative action of the heterotic string ($b=0$) 
encodes the function ${\cal F}$, and thereby the entire action. In particular, 
the result for the bosonic string ($a=b$) follows or can be 
reconstructed from the heterotic result. 
Moreover, in the heterotic case we note the formal ``gauge invariance''
 \be
  \veccomp\rightarrow \veccomp+ \eta\bar{P}\Lambda,
 \ee
 under which the action (\ref{eq:I1d+1currentsproj}) (for $b=0$) is invariant.

We may also express the result in terms of the $\O(d+1,d+1)$ coset currents (\ref{eq:Pd+1}).
In the basis~\eqref{eq:Pab}, the four-derivative action~\eqref{eq:I1d+1currentsproj} takes the form
\begin{equation} \label{eq:I1d+1proj}
 \begin{aligned}
  \widetilde{I}_{1} & = \frac{\alpha'}{4}\int\d^{3}x\,\sqrt{-g}\,e^{-2\Phi}\,\Bigg\{\\
  &+\frac{a}{4}\bigg[\Tr{{\sf P}_{\mu}{\sf P}^{{\rm t}\,\mu}{\sf P}_{\nu}{\sf P}^{{\rm t}\,\nu}}-\Tr{{\sf P}_{\mu}^{\rm t}{\sf P}^{\mu}{\sf P}_{\nu}^{\rm t}{\sf P}^{\nu}}-\Tr{{\sf P}_{\mu}{\sf P}^{\rm t}_{\nu}{\sf P}^{\mu}{\sf P}^{{\rm t}\,\nu}}-\Tr{{\sf P}_{\mu}{\sf P}^{\rm t}_{\nu}}\Tr{{\sf P}^{\mu}{\sf P}^{{\rm t}\,\nu}} \\
  &\quad\ \,\quad+\frac{1}{2}\,\Tr{{\sf P}_{\mu}{\sf P}^{{\rm t}\,\mu}}\Tr{{\sf P}_{\nu}{\sf P}^{{\rm t}\,\nu}} +4\,\left({\sf P}_{\mu}{\sf P}^{\rm t}_{\nu}{\sf P}^{\mu}{\sf P}^{{\rm t}\,\nu}\right)^{00}-4\,\left({\sf P}_{\mu}{\sf P}^{{\rm t}\,\mu}{\sf P}_{\nu}{\sf P}^{{\rm t}\,\nu}\right)^{00}+4\,\left({\sf P}_{\mu}{\sf P}^{\rm t}_{\nu}{\sf P}^{\nu}{\sf P}^{{\rm t}\,\mu}\right)^{00}\\
  &\quad\ \,\quad+4\,\Tr{{\sf P}_{\mu}{\sf P}^{\rm t}_{\nu}}\left({\sf P}^{\mu}{\sf P}^{{\rm t}\,\nu}\right)^{00}-2\,\Tr{{\sf P}_{\mu}{\sf P}^{{\rm t}\,\mu}}\left({\sf P}_{\nu}{\sf P}^{{\rm t}\,\nu}\right)^{00}-8\,\left({\sf P}_{\mu}{\sf P}^{\rm t}_{\nu}\right)^{00}\left({\sf P}^{\mu}{\sf P}^{{\rm t}\,\nu}\right)^{00}\\
  &\quad\ \,\quad+4\,\left({\sf P}_{\mu}{\sf P}^{{\rm t}\,\mu}\right)^{00}\left({\sf P}_{\nu}{\sf P}^{{\rm t}\,\nu}\right)^{00}\bigg]\\
  &+\frac{b}{4}\,\bigg[-\Tr{{\sf P}_{\mu}{\sf P}^{{\rm t}\,\mu}{\sf P}_{\nu}{\sf P}^{{\rm t}\,\nu}}+\Tr{{\sf P}_{\mu}^{\rm t}{\sf P}^{\mu}{\sf P}_{\nu}^{\rm t}{\sf P}^{\nu}}-\Tr{{\sf P}_{\mu}{\sf P}^{\rm t}_{\nu}{\sf P}^{\mu}{\sf P}^{{\rm t}\,\nu}}-\Tr{{\sf P}_{\mu}{\sf P}^{\rm t}_{\nu}}\Tr{{\sf P}^{\mu}{\sf P}^{{\rm t}\,\nu}}\\
  &\quad\ \,\quad+\frac{1}{2}\,\Tr{{\sf P}_{\mu}{\sf P}^{{\rm t}\,\mu}}\Tr{{\sf P}_{\nu}{\sf P}^{{\rm t}\,\nu}}+4\,\left({\sf P}^{\rm t}_{\mu}{\sf P}_{\nu}{\sf P}^{{\rm t}\,\mu}{\sf P}^{\nu}\right)^{\bar 0\bar 0}-4\,\left({\sf P}^{\rm t}_{\mu}{\sf P}^{\mu}{\sf P}^{\rm t}_{\nu}{\sf P}^{\nu}\right)^{\bar 0\bar 0}+4\,\left({\sf P}^{\rm t}_{\mu}{\sf P}_{\nu}{\sf P}^{{\rm t}\,\nu}{\sf P}^{\mu}\right)^{\bar 0\bar 0} \\
  &\quad\ \,\quad +4\,\Tr{{\sf P}_{\mu}{\sf P}^{\rm t}_{\nu}}\left({\sf P}^{{\rm t}\,\mu}{\sf P}^{\nu}\right)^{\bar 0\bar 0}-2\,\Tr{{\sf P}_{\mu}{\sf P}^{{\rm t}\,\mu}}\left({\sf P}^{{\rm t}}_{\nu}{\sf P}^{\nu}\right)^{\bar 0\bar 0}-8\,\left({\sf P}^{\rm t}_{\mu}{\sf P}_{\nu}\right)^{\bar 0 \bar 0}\left({\sf P}^{{\rm t}\,\mu}{\sf P}^{\nu}\right)^{\bar 0\bar 0}\\
  &\quad\ \,\quad+4\,\left({\sf P}^{{\rm t}}_{\mu}{\sf P}^{\mu}\right)^{\bar 0\bar 0}\left({\sf P}^{\rm t}_{\nu}{\sf P}^{\nu}\right)^{\bar 0\bar 0}\bigg]\Bigg\}.
 \end{aligned}
\end{equation}
Here, the `$^0$' and `$^{\bar 0}$' components are defined by contracting out $\O(d+1)\times\O(d+1)$ vectors as
\be
{\sf P}_\mu^{0\bar a} = e^{-\Phi} \, \widetilde{\veccomp}_a {\sf P}_\mu^{a\bar a}
\;,\qquad
{\sf P}_\mu^{a\bar 0} =  {\sf P}_\mu^{a\bar a}\, \widetilde{\veccomp}_{\bar a} \, e^{-\Phi}
\;,
\label{eq:PP00}
\ee
with $\widetilde{\veccomp}_a  = \veccomp^M {\cal V}_{Ma}$, $\widetilde{\veccomp}_{\bar a}  = \veccomp^M {\cal V}_{M\bar a}$.
Spelling out eq.~\eqref{eq:PP00} brings the action into manifestly $H$ gauge invariant form. In these notations, ${\sf P}$ and  ${\sf P}^{\rm t}$ are interchanged under the $\mathbb{Z}_{2}$ action.

More compactly, and using the notations of refs.~\cite{Bossard:2016zdx,Bossard:2017wum},
the result (\ref{eq:I1d+1proj}) can be rewritten in the form
\begin{equation}
 \begin{aligned}
  \widetilde{I}_{1} = \frac{\alpha'}{4}\int\d^{3}x\,\sqrt{-g}\,e^{-2\Phi}\,&\Bigg[\frac{a}{4}\,F_{abcd}\left(({\sf P}_{\mu}{\sf P}^{{\rm t}\,\mu})^{ab}({\sf P}_{\nu}{\sf P}^{{\rm t}\,\nu})^{cd}-2\,({\sf P}_{\mu}{\sf P}^{\rm t}_{\nu})^{ab}({\sf P}^{\mu}{\sf P}^{{\rm t}\,\nu})^{cd}\right)\\
  &+\frac{b}{4}{\bar F}_{\bar a\bar b\bar c\bar d}\left(({\sf P}^{\rm t}_{\mu}{\sf P}^{\mu})^{\bar a\bar b}({\sf P}^{\rm t}_{\nu}{\sf P}^{\nu})^{\bar c\bar d}-2\,({\sf P}^{\rm t}_{\mu}{\sf P}_{\nu})^{\bar a\bar b}({\sf P}^{{\rm t}\,\mu}{{}\sf P}^{\nu})^{\bar c\bar d}\right)\Bigg],
 \end{aligned}
 \label{eq:I1d+1PP}
\end{equation}
with
\begin{equation}\label{FTensor} 
  F_{abcd} = \frac{3}{2}\,\delta_{(ab}\delta_{cd)} - 6\,\delta_{0(a}\delta_{bc}\delta_{d)0} +4\,\delta_{0a}\delta_{0b}\delta_{0c}\delta_{0d},
\end{equation}
and ${\bar F}_{\bar a\bar b\bar c\bar d}$ defined in the same way, exchanging unbared indices for bared ones. 
In the case of the heterotic supergravity, $(a,b)=(-\alpha',0)$, eq.~\eqref{eq:I1d+1PP} consistently reproduces the weak 
coupling limit of the U-duality invariant modular integrals conjectured to describe the exact four-derivative couplings, \textit{c.f.}~ref.~\cite{Bossard:2016zdx,Bossard:2017wum,Obers:2000ta}. (See in particular eqs.~(4.16) and (4.34) of ref.~\cite{Bossard:2017wum}.\footnote{We thank Guillaume Bossard for helpful
explanations on this relation.})

\subsection{Comments on symmetry group} 

We close this section with some general remarks on the symmetry group after inclusion of the 
first $\alpha'$ correction, confirming our above conclusion that the continuous $\O(8,8)$ duality group is broken 
to its `geometric subgroup' and making some general remarks about the issue of discrete symmetries. 

We begin by determining  the manifest symmetry group. Since the above formulation is manifestly $\O(8,8)$ invariant 
if we declare the compensator $\veccomp^{\cal M}$ to transform as a vector it follows that the actual, or at least manifest,  
symmetry group is given by the invariance group of $\veccomp^{\cal M}=(0,1,0)$. 
Recalling  the index split ${\cal M}=(M,+,-)$, an $\O(8,8)$ matrix reads 
 \be
  L_{{\cal M}}{}^{\cal N} = \begin{pmatrix} L_{M}{}^{N} & L_{M}{}^+ & L_M{}^- \\
  L_+{}^{N} & L_+{}^+ & L_+{}^- \\
  L_-{}^{N} & L_-{}^+ & L_-{}^- \end{pmatrix} \; \in \; \O(8,8)\;, 
 \ee 
and is subject to 
 \be\label{groupProperties}
  L_{{\cal M}}{}^{\cal K} \,\eta_{{\cal K}{\cal L}}\, L_{{\cal N}}{}^{\cal L} = \eta_{{\cal M}{\cal N}}\;. 
 \ee
The condition that $\veccomp^{\cal M}$ is invariant, \textit{i.e.} that $\veccomp'^M=0$, $\veccomp_+'=0$ and $\veccomp_-'=1$, 
is then quickly seen to imply 
 \be
  L_M{}^- = 0\;, \qquad L_+{}^- =0 \;, \qquad L_-{}^{-}=1\;. 
 \ee
Using this in eq.~\eqref{groupProperties} yields furthermore 
 \be
 \begin{split}
  L_M{}^N \in \O(7,7)\;, \quad L_+{}^M&=0\;, \quad 
  (L^{-1})_N{}^{M} L_-{}^{N} = -\eta^{MN} L_N{}^{+}\;,  \\ 
  L_+{}^+&=1\;, \qquad 
  L_-{}^{+} = -\frac{1}{2} L_-{}^{M} \eta_{MN} L_-{}^{N}\;. 
 \end{split}
 \ee 
We infer that a  general  $\O(8,8)$ matrix leaving $\veccomp$ invariant is parametrized in terms of  a general $\O(7,7)$ matrix 
$L_M{}^N$ and a vector $c_{M}$:
\begin{equation}
  L_{\cal M}{}^{\cal N} = \begin{pmatrix}
                            L_{M}{}^{N} & L_{M}{}^{K}c_{K} & 0 \\
                            0 & 1 & 0 \\
                            -c^{N} & -\frac{1}{2}\,c_{K}c^{K} & 1
                          \end{pmatrix}\,. 
\end{equation}
One may verify that the $c_{M}$, for $L_M{}^N=\delta_M{}^N$, act as 
 \be
  \xi_M' = \xi_M+ c_M\;, 
 \ee
and thus precisely parameterize the  constant shifts of 
the scalars dual to vectors.  As expected, we recovered  precisely the `geometric subgroup' of $\O(8,8)$
consisting of the semi-direct product of $\O(7,7)$ and $14$-dimensional translations.

We thus conclude  that among the continuous transformations there is no enhancement to 
the full duality group, but this still leaves open the fate of the discrete group $\O(8,8,\mathbb{Z})$,\,which is conjectured to be present in the full string/M-theory. 
Ultimately we see no evidence of this true U-duality in the supergravity limit considered here 
once $\alpha'$ corrections are switched on, but we will briefly discuss  some general 
features.  

We first observe that the scaling symmetry denoted (Dilaton$^3$)  above, which acts as $\Phi\rightarrow \Phi+\lambda$, 
$\lambda\in \mathbb{R}$, 
on the dilaton and is hence manifestly broken due to the explicit $e^{-2\Phi}$ prefactors in the action at order $\alpha'$, 
trivializes for the discrete subgroup. In fact, these transformations are embedded into $\O(8,8)$ as 
 \be\label{scalingMatrix}
    \lambda\;  \mapsto \; L_{(\lambda)} = \begin{pmatrix} {\bf 1}_{7} & 0 & 0 \\
    0 & e^{\lambda} & 0 \\
    0 & 0 & e^{-\lambda}  \end{pmatrix}  \ \in \ \O(8,8;\mathbb{R}) \;, 
   \ee 
but the requirement that this transformation  actually belongs to $\O(8,8,\mathbb{Z})$, \textit{i.e.} that all 
its matrix entries are integers then implies that $e^{\lambda}=e^{-\lambda}= 1$ or $\lambda=0$, reducing (Dilaton$^3$) to the trivial group.
Thus, by itself the absence of the continuous  scaling symmetry at order $\alpha'$ is \textit{not} in conflict with an 
$\O(8,8,\mathbb{Z})$ symmetry.

Is then the 
$\O(8,8,\mathbb{Z})$ perhaps realized in supergravity after all? 
While it is very difficult to exclude the possibility that certain discrete symmetries are realized 
in a hidden fashion, we will argue now that there is no evidence for any enhancement to the 
discrete U-duality group in supergravity at order $\alpha'$. 
 To this end one may use, as discussed by Sen in ref.~\cite{Sen:1994wr}, that $\O(8,8,\mathbb{Z})$ 
 can be generated by 
 $\O(7,7,\mathbb{Z})$ and the S-duality group $\SL(2,\mathbb{Z})$. 
 As the invariance  under $\O(7,7,\mathbb{Z})$ is manifest, it is then sufficient to study the invariance under $\SL(2,\mathbb{Z})$, 
which consists of matrices
  \begin{equation}\label{SL(2,Z)rules} 
    \begin{pmatrix}
      c & d\\
      e & f
    \end{pmatrix}, \quad c,d,e,f\in\mathbb{Z}\,, \; \;{\rm where}\quad  cf-de=1\;. 
  \end{equation}
These are embedded  into 
$\O(8,8,\mathbb{Z})$ as follows:
  \begin{equation}
    L_{\cal M}{}^{\cal N} = \begin{pmatrix}
                                  c & 0 & 0 & 0 & d & 0 \\
                                  0 & \mathds{1}_{d-1} & 0 & 0 & 0 & 0 \\
                                  0 & 0 & f & 0 & 0 & -\,e \\
                                  0 & 0 & 0 & \mathds{1}_{d-1} & 0 & 0 \\
                                  e & 0 & 0 & 0 & f & 0 \\
                                  0 & 0 & -d & 0 & 0 & c
                                 \end{pmatrix}.
  \end{equation}
  Under such transformations, a generic term $\left(\veccomp X\veccomp\right)$ of the action~\eqref{eq:I1d+1currentsproj} transforms as
  \begin{equation}  \label{eq:discretetransfo}
    \left(\veccomp X\veccomp\right)=X_{++} \longrightarrow f^{2}\,X_{++}+2\,ef\,X_{(+1)}+e^{2}\,X_{11}.
  \end{equation}
Thus,  each term is independently invariant only if $c=f=1$ and $e=0$, leaving only integer shifts~\eqref{eq:xishift}. 
Therefore, non-trivial $\SL(2,\mathbb{Z})$ transformations can only  leave the action invariant if the different anomalous terms cancel each other. Since the traces in eq.~\eqref{eq:I1d+1currentsproj} are invariant, and the transformation rule~\eqref{eq:discretetransfo} cannot generate trace factors,  transformations of terms featuring traces can only cancel among  themselves.
Considering  the pure trace terms in the first and sixth lines of eq.~\eqref{eq:I1d+1currentsproj}, one infers that due to 
the global dilaton prefactor, these terms are only invariant under transformations that leave the dilaton invariant, and so with eq.~\eqref{eq:genmetd+1},
  \begin{equation}
    e^{2\Phi}=\left(\veccomp\M\veccomp\right) \longrightarrow \left(f+e\,\xi_{1}\right)^{2}e^{2\Phi}+e^{2}\,\H_{11}\,,
  \end{equation}
implying as before  $f=1$ and $e=0$, and hence $c=1$ by the $\SL(2,\mathbb{Z})$ constraint (\ref{SL(2,Z)rules}). 
Even discarding dilaton prefactors we have found no evidence of any hidden discrete symmetries, 
supporting the above conclusion that only the continuous geometric subgroup of $\O(8,8)$ 
is a genuine symmetry.

\section{A massive deformation} \label{sec:Bfield}
So far the three form field-strength $H_{\mu\nu\rho}$ has been set to zero. We now integrate out the $B$ field explicitly and explore the deformations induced by a non-vanishing three-form flux. At the two-derivative level, this results in a topological mass for the vectors and a potential for the dilaton~\cite{Kaloper:1993fg}. We review this massive deformation and show how it fits in the more general framework of gauged supergravity. We then extend the analysis to the four-derivative corrections. In particular, we show that the resulting massive deformation induces Chern-Simons terms for composite connections featuring an enhancement to the full $\O(d+1,d+1)$ to first order in the mass parameter.

  \subsection{Two-derivative action}  \label{sec:Bfieldtwoder}
  We first consider the two-derivative action~\eqref{eq:I0}, that we rewrite in the string frame as
  \begin{equation}
    I_{0}=\int\d^{3}x\,\sqrt{-g}\,e^{-\Phi}\left({\cal L}^{(0)}\left(g_{\mu\nu},\Phi,\H,\A_{\mu}{}^{M}\right)-\frac{1}{12}\,H_{\mu\nu\rho}H^{\mu\nu\rho}\right),
  \end{equation}
  with $H_{\mu\nu\rho}=3\,\partial_{[\mu}B_{\nu\rho]}+\,\Omega^{\rm (MS)}_{\mu\nu\rho}$ and $\Omega^{\rm (MS)}_{\mu\nu\rho}=-(3/2)\,\A_{[\mu}{}^{M}\F_{\nu\rho]\,M}$ the abelian Chern-Simons deformation of the three-form field-strength~\cite{Maharana:1992my}. In three dimensions, we can rewrite the field-strength as\footnote{The global factor is chosen so that $h=\epsilon^{\mu\nu\rho}H_{\mu\nu\rho}$, with $g_{\mu\nu}$ of signature $(-1,1,1)$.}
  \begin{equation}  \label{eq:H3d}
    H_{\mu\nu\rho} = -\frac{1}{6}\,h\,\epsilon_{\mu\nu\rho},
  \end{equation}
  and $I_{0}$ becomes
  \begin{equation}  \label{eq:I0h}
    I_{0}=\int\d^{3}x\,\sqrt{-g}\,e^{-\Phi}\left({\cal L}^{(0)}\left(g_{\mu\nu},\Phi,\H,\A_{\mu}{}^{M}\right)+\frac{1}{72}\,h^{2}\right).
  \end{equation}
  In the following, we do not specify explicitly the fields on which ${\cal L}^{(0)}$ depends. To dualise the degrees of freedom related to $B_{\mu\nu}$, we introduce an auxiliary field $f$ and consider the action
  \begin{equation}  \label{eq:primeI0}
    I'_{0}=\int\d^{3}x\,\sqrt{-g}\,e^{-\Phi}\left({\cal L}^{(0)}-\frac{1}{2}\,f^{2}-\frac{1}{6}\,fh\right).
  \end{equation}
  Taking its variation with respect to $f$, we get the algebraic equation of motion
  \begin{equation}  \label{eq:eomf0}
    f=-\frac{1}{6}\,h,
  \end{equation}
  which can be used in eq.~\eqref{eq:primeI0} to get back eq.~\eqref{eq:I0h}, hence the equivalence between $I_{0}$ and $I'_{0}$. We now consider $I'_{0}$, which gives the following equation of motion after varying with respect to $B_{\mu\nu}$:
  \begin{equation}    \label{eq:eomB0}
    \nabla_{\mu}\left(e^{-\Phi}f\right) = 0 \quad \implies \quad f = m\,e^{\Phi},\ m\in\mathbb{R}.
  \end{equation}
  Unlike eq.~\eqref{eq:eomf0}, this equation is not algebraic and, in general, could not simply be inserted back into the action. One needs to work at the level of the equations of motion. It can however be checked that, in the particular case we are considering here, it is consistent to use eq.~\eqref{eq:eomB0} directly in the action~\eqref{eq:primeI0} to get
  \begin{equation}  \label{eq:secondI0}
    I''_{0}=\int\d^{3}x\left[\sqrt{-g}\left(e^{-\Phi}\,{\cal L}^{(0)}-\frac{1}{2}\,m^{2}\,e^{\Phi}\right) -\frac{1}{6}\,m\,\varepsilon^{\mu\nu\rho}\,\Omega^{\rm(MS)}_{\mu\nu\rho}\right],
  \end{equation}
  where we ignored a total derivative. $I''_{0}$ is equivalent to $I_{0}$, with the degrees of freedom of the two-form $B_{\mu\nu}$ dualised into $m$. For $m=0$, we recover the action we started with in sec.~\ref{sec:dualityenhancement}.

  Let us move to the Einstein frame to discuss further the properties of the action:
  \begin{equation}  \label{eq:secondI0einstein}
    \begin{aligned}
      I''_{0}=\int\d^{3}x\bigg[\sqrt{-g_{\rm E}}\,\bigg(&R_{\rm E}-\partial_{\mu}\Phi\,\partial^{\mu}\Phi+\frac{1}{8}\,\Tr{\partial_{\mu}{\cal H}^{-1}\,\partial^{\mu}{\cal H}}-\frac{1}{4}\,e^{-2\Phi}\,\F_{\mu\nu}{}^{M}\H_{MN}\F^{\mu\nu\,N}\\
                  &-\frac{1}{2}\,m^{2}\,e^{4\Phi}\bigg)+\frac{1}{4}\,m\,\varepsilon^{\mu\nu\rho}\,\A_{\mu}{}^{M}\F_{\nu\rho\,M}\bigg].
    \end{aligned}
  \end{equation}
  For $m\neq0$, the $\O(d+1,d+1)$ symmetry of the action~\eqref{eq:I0d+1} is broken down to \Odd and constant shifts in $\xi_{M}$ (the scaling symmetry~\eqref{eq:phi3} is broken). The term quadratic in $m$ acts as a potential, and the vectors $\A_{\mu}{}^{M}$ acquire a topological mass proportional to $m$, leading to a topologically massive Yang-Mills theory~\cite{Schonfeld:1980kb,Deser:1982vy} for ${\rm U}(1)^{d}$, with equations of motion
  \begin{equation}  \label{eq:dual2derm}
    \nabla_{\mu}\Big(e^{-2\Phi}\,\F^{\mu\nu\,N}\H_{NM}-m\,\epsilon^{\mu\nu\rho}\,\A_{\rho\,M}\Big) = 0.
  \end{equation}
  The Chern-Simons coupling prevents us from dualizing the vectors as done in sec.~\ref{sec:dualityenhancement}. We can however rewrite the Yang-Mills gauging of eq.~\eqref{eq:secondI0einstein} as a pure Chern-Simons type gauging with gauge group ${\rm U}(1)^{d}\ltimes T_d$ using the on-shell equivalence of ref.~\cite{Nicolai:2003bp}, where $T_d$ is a $d$-dimensional translation group. Consider the action
  \begin{equation} \label{eq:secondI0tildeeinstein}
    \begin{aligned}
      \widetilde{I}''_{0}=\int\d^{3}x\bigg[&\sqrt{-g_{\rm E}}\,\Big(R_{\rm E}+\frac{1}{8}\,\Tr{\partial_{\mu}\M^{-1}\,\partial^{\mu}\M}-\frac{1}{2}\,m^{2}\,e^{2\Phi}\,\A_{\mu}{}^{M}\H_{MN}\A^{\mu\,N}\\
      &-m\,e^{2\Phi}\,\A_{\mu}{}^{M}\H_{MN}\partial^{\mu}\xi^{N}-\frac{1}{2}\,m^{2}\,e^{4\Phi}\Big)-\frac{1}{4}\,m\,\varepsilon^{\mu\nu\rho}\,\A_{\mu}{}^{M}\F_{\nu\rho\,M}\bigg],
    \end{aligned}
   \end{equation}
   The induced equations of motion for the vectors are
   \begin{equation} \label{eq:dual2dermtheta}
    \F_{\mu\nu}{}^{M} = e^{2\Phi}\,\epsilon_{\mu\nu\rho}\left(\partial^{\rho}\xi^{N}+m\,\A^{\rho\,N}\right)\H_{N}{}^{M},
   \end{equation}
   which imply eq.~\eqref{eq:dual2derm}. It can be checked similarly that all the equations of motion of $\widetilde{I}''_{0}$ are identical on-shell to those of $I''_{0}$, upon systematically eliminating $\partial_{\mu}\xi_{M}$ using eq.~\eqref{eq:dual2dermtheta}. Thus, $I''_{0}$ and $\widetilde{I}''_{0}$ are equivalent.

   The action~\eqref{eq:secondI0tildeeinstein} can be nicely rewritten using the embedding tensor formalism~\cite{Nicolai:2000sc,deWit:2002vt} of three-dimensional half-maximal gauged supergravity~\cite{Nicolai:2001ac,deWit:2003ja} (see also ref.~\cite{Eloy:2021qol} for a review and the notations used here). The bosonic part of the action describes the dynamics of the metric $g_{\rm E\,\mu\nu}$, scalars $\M_{\cal MN}\in\O(d+1,d+1)$ and vectors $A_{\mu}{}^{\cal [MN]}$ via the action
  \begin{equation}
    \int\d^{3}x\left[\sqrt{-g_{\rm E}}\,\Big(R_{\rm E}+\frac{1}{8}\,\Tr{D_{\mu}\M^{-1}\,D^{\mu}\M}- V\Big)+{\cal L}_{\rm CS}\right].
  \end{equation}
  The covariant derivative on $\M_{\cal MN}$ is
  \begin{equation}
    D_{\mu}\M_{\cal MN}=\partial_{\mu}\M_{\cal MN} + 4\,A_{\mu}{}^{\cal PQ}\Theta_{\cal PQ\vert (M}{}^{\cal K}\M_{\cal N)K},
  \end{equation}
  with the gauging given by the embedding tensor
  \begin{equation}
    \Theta_{\cal M N\vert P Q} = \frac{1}{2}\left(\eta_{\cal M[P}\,\theta_{\cal Q]N}-\eta_{\cal N[P}\,\theta_{\cal Q]M}\right),
  \end{equation}
  where $\theta_{\cal MN}$ is symmetric. In full generality, the embedding tensor contains more representations that we will not need here. The vectors are described by a Chern-Simons term of the form
  \begin{equation}
    {\cal L}_{\rm CS}= -\varepsilon^{\,\mu\nu\rho}\,\Theta_{\cal M N\vert P Q}\,A_{\mu}{}^{\cal MN}\left(\partial_{\nu}\,A_{\rho}{}^{\cal PQ}  + \frac{1}{3}\, \Theta_{\cal RS\vert UV}\,f^{\cal PQ,RS}{}_{\cal XY}\, A_{\nu}{}^{\cal UV} A_{\rho}{}^{\cal XY} \right),
  \end{equation}
  with $f^{\cal MN,PQ}{}_{\cal KL} = 4\,\delta_{[\cal K}{}^{[\cal M}\eta^{\cal N][P}\delta_{\cal L]}{}^{\cal Q]}$ the structure constants of $\mathfrak{so}(d+1,d+1)$. Finally, the potential is given by
  \begin{equation}
    V=\frac{1}{8}\,\theta_{\cal MN}\theta_{\cal PQ}\left(2\,\M^{\cal MP}\M^{\cal NQ}-2\,\eta^{\cal MP}\eta^{\cal NQ}-\M^{\cal MN}\M^{\cal PQ}\right).
  \end{equation}
  The action~\eqref{eq:secondI0tildeeinstein} then results from the restriction to an embedding tensor with only non-vanishing component $\theta_{--}$. More precisely, a formally $\O(d+1,d+1)$-covariant form of $I''_{0}$ is given by
   \begin{equation}
    \begin{aligned}
      \widetilde{I}''_{0}=\int\d^{3}x\bigg[&-\varepsilon^{\,\mu\nu\rho}\,\theta_{\cal MN}\,A_{\mu\,{\cal P}}{}^{\cal N}\partial_{\nu}\,A_{\rho}{}^{\cal PM}+\sqrt{-g_{\rm E}}\,\bigg(R_{\rm E}+\frac{1}{8}\,\Tr{D_{\mu}\M^{-1}\,D^{\mu}\M} \\
      &-\frac{1}{8}\,\theta_{\cal MN}\theta_{\cal PQ}\left(2\,\M^{\cal MP}\M^{\cal NQ}-2\,\eta^{\cal MP}\eta^{\cal NQ}-\M^{\cal MN}\M^{\cal PQ}\right)\bigg)\bigg],
    \end{aligned}
   \end{equation}
   with the specific parametrization
   \begin{equation} \label{eq:embeddingm}
     A_{\mu}{}^{M-}=\frac{1}{2}\,\A_{\mu}{}^{M},\quad \theta_{--}=2\,m.
   \end{equation}
   The covariant derivative is then given by $D_{\mu}\xi_{M}=\partial_{\mu}\xi_{M}+m\,\A_{\mu\,M}$, $D_{\mu}\Phi=\partial_{\mu}\Phi$ and $D_{\mu}\H_{MN}=\partial_{\mu}\H_{MN}$. The symmetry breaking induced by $m\neq0$ is now translated into the choice of embedding tensor with $\theta_{--}$ as the only non-vanishing component, that breaks $\O(d+1,d+1)$ to \Odd and shifts in $\xi_{M}$.

  \subsection{Four-derivative action}
  The 2-form degrees of freedom can be integrated out in the four-derivative action using the same procedure as the one used in sec.~\ref{sec:Bfieldtwoder}. With eq.~\eqref{eq:H3d}, the action (7.16) of ref.~\cite{Eloy:2020dko} is given by
  \begin{equation}
   \widetilde{I} = I + \int\d^{3}x\,\sqrt{-g}\,e^{-\Phi}\left[\frac{1}{72}\,h^{2}+\alpha'\left(-\frac{1}{6}\,\mathfrak{a}\,h+\frac{1}{36}\,\mathfrak{b}\,h^{2}+\frac{1}{6^{4}}\,\mathfrak{c}\,h^{4}\right)\right],
  \end{equation}
  where $I$ is the action~\eqref{eq:Ifull} and\footnote{Remember that $a$ and $b$ are of order $\alpha'$.}
  \begin{equation}
    \begin{aligned}
      \mathfrak{a} &= \epsilon^{\mu\nu\rho}\bigg[-\frac{a+b}{8\,\alpha'}\left(\frac{2}{3}\,\Omega_{\mu\nu\rho}^{\rm (GS)}-\frac{1}{2}\,\F_{\mu\sigma}{}^{M}\left(\H\partial_{\nu}\H^{-1}\right)_{M}{}^{N}\F_{\rho}{}^{\sigma}{}_{N}\right)\\
      &\qquad\qquad+\frac{a-b}{4\,\alpha'}\left(\Omega_{\mu\nu\rho}^{(\omega)}+\frac{1}{4}\,\F_{\mu\sigma}{}^{M}\partial^{\sigma}\H_{MN}\F_{\nu\rho}{}^{N}\right)\bigg],\\
      \mathfrak{b} &=-\frac{a+b}{8\,\alpha'}\left(R-\frac{1}{4}\,\Tr{\partial_{\mu}\H\partial^{\mu}\H^{-1}}+\frac{3}{4}\,\F_{\mu\nu}{}^{M}\H_{MN}\F^{\mu\nu\,N}\right)+\frac{a-b}{4\,\alpha'}\,\frac{1}{8}\,\F_{\mu\nu}{}^{M}\F^{\mu\nu}{}_{M},\\ 
      \mathfrak{c}&=\frac{a+b}{8\,\alpha'}\,\frac{5}{4}.
    \end{aligned}
  \end{equation}
  $\Omega^{(\omega)}_{\mu\nu\rho}$ is the gravitational Chern-Simons form of heterotic supergravity, and $\Omega^{\rm (GS)}_{\mu\nu\rho}$ is the three-form needed in the Green-Schwarz type mechanism of ref.~\cite{Eloy:2019hnl}, which satisfies\footnote{Normalised as in eq.~(14) of ref.~\cite{Eloy:2019hnl}, \textit{i.e.} $\widetilde{H}_{\mu\nu\rho} = H_{\mu\nu\rho}+\frac{a+b}{2}\,\Omega^{\rm (GS)}_{\mu\nu\rho}$.}
  \begin{equation}
    4\,\partial_{[\mu}\Omega^{\rm (GS)}_{\nu\rho\sigma]}=\frac{3}{8}\,\Tr{\partial_{[\mu}\H\partial_{\nu}\H^{-1}\partial_{\rho}\H\partial_{\sigma]}\H^{-1}\H\eta}.
  \end{equation}

  As previously, we can equivalently write $\widetilde{I}$ as
  \begin{equation}  \label{eq:primeI1}
    \widetilde{I}'=I+\int\d^{3}x\,\sqrt{-g}\,e^{-\Phi}\left[-\frac{1}{2}\,f^{2}-\frac{1}{6}\,fh+\alpha'\left(\mathfrak{a}\,f+\mathfrak{b}\,f^{2}+\mathfrak{c}\,f^{4}\right)\right],
  \end{equation}
  with an auxiliary field $f$. The equations of motion for $f$ are
  \begin{equation}
    f=-\frac{1}{6}\,h+\alpha'\left(\mathfrak{a}+2\,\mathfrak{b}\,f+4\,\mathfrak{c}\,f^{3}\right),
  \end{equation}
  and can be solved perturbatively in $\alpha'$:
  \begin{equation}  \label{eq:eomf1}
    f=f^{(0)} + \alpha'\,f^{(1)},\ {\rm with}\ 
    \begin{cases}
      \displaystyle f^{(0)}=-\frac{1}{6}\,h,\\
      \displaystyle f^{(1)}=\mathfrak{a}-\frac{1}{3}\,\mathfrak{b}\,h-\frac{1}{54}\,\mathfrak{c}\,h^{3}.
    \end{cases}
  \end{equation}
  As in the two-derivative case, eq.~\eqref{eq:primeI1} with eq.~\eqref{eq:eomf1} gives $\widetilde{I}$. The equations of motion for $B_{\mu\nu}$ are unchanged and given by eq.~\eqref{eq:eomB0} and, again, it is consistent to use them directly in the action, leading to 
  \begin{equation} \label{eq:I1second}
    \widetilde{I}''=I+\int\d^{3}x\,\sqrt{-g}\left[-\frac{1}{6}\,m\,\epsilon^{\mu\nu\rho}\,\Omega^{\rm(MS)}_{\mu\nu\rho}-\frac{1}{2}\,m^{2}\,e^{\Phi}+\alpha'\left(\mathfrak{a}\,m\!+\!\mathfrak{b}\,m^{2}\,e^{\Phi}+\mathfrak{c}\,m^{4}\,e^{3\Phi}\right)\right].
  \end{equation}
  As a by product, observe that we can safely consider the case $m=0$ and recover the actions considered in sec.~\ref{sec:I0} and \ref{sec:I1}.

  Writing this action in terms of $\O(d+1,d+1)$ fields and an embedding tensor breaking the symmetry to $\Odd\times\O(1,1)$, as we did for the two-derivative action in sec.~\ref{sec:Bfieldtwoder}, requires to reproduce the analysis of sec.~\ref{sec:I1} with modified rules for the field redefinitions (given the modified two-derivative equations of motion) and with the additional terms in eq.~\eqref{eq:I1second}, as detailed in app.~\ref{app:4derBfield}. All computations done, we get
\begin{equation} \label{eq:I1secondd+1currents}
 \begin{aligned}
  \widetilde{I}''_{1} &= \widehat{I}_{1}+\int\d^{3}x\,\Bigg\{\sqrt{-g_{\rm E}}\,\frac{a+b}{8}\,\bigg(4\,m^{2}\,e^{2\Phi}\,\partial_{\mu}\Phi\partial^{\mu}\Phi+m^{4}\,e^{6\Phi}\bigg)\\
  &+m\,\varepsilon^{\mu\nu\rho}\bigg[-\frac{a+b}{8}\bigg(\frac{2}{3}\,\Omega_{\mu\nu\rho}^{(\rm GS)}-\frac{1}{2}\,e^{2\Phi}\,D_{\mu}\xi_{M}\left(\H^{-1}\partial_{\nu}\H\right)^{M}{}_{N}D_{\rho}\xi^{N}\bigg)+\frac{a-b}{4}\,\Omega_{\mu\nu\rho}^{(\omega_{\rm E})}\bigg]\Bigg\},
 \end{aligned}
\end{equation}
where $\widehat{I}_{1}$ is given by the action~\eqref{eq:I1d+1currentsproj} with all currents covariantized: $\widehat{\J}_{\mu}= D_{\mu}\M\M^{-1}$. Observe that the Green-Schwarz type mechanism of ref.~\cite{Eloy:2019hnl} generates, once restricted to three dimensions, a Chern-Simons term based on composite gauge fields. The properties of this term are best displayed using the frame formalism of sec.~\ref{sec:dualityenhancement}. In ref.~\cite{Eloy:2019hnl}, the Green-Schwarz three-from $\Omega_{\mu\nu\rho}^{\rm (GS)}$ has been written as a Chern-Simons form for $\O(d)\times O(d)$ composite gauge fields:
 \begin{equation} \label{eq:GSdasframe}
   \Omega_{\mu\nu\rho}^{\rm (GS)} = \frac{3}{2}\,{\rm CS}_{\mu\nu\rho}\left(Q\right)=\frac{3}{2}\,\Tr{Q_{[\mu}\partial_{\nu}Q_{\rho]}\delta\eta + \frac{2}{3}\,Q_{[\mu}Q_{\nu}Q_{\rho]}\delta\eta},
 \end{equation}
 with $Q_{\mu\,A}{}^{B}$ defined in eq.~\eqref{eq:Qd+1} and $\delta_{AB}$ the identity matrix.\footnote{The $\delta\eta$ in eq.~\eqref{eq:GSdasframe} is needed to reproduce the right relative sign in eq.~(34) of ref.~\cite{Eloy:2019hnl}.} Generalizing this definition to the covariantization $\widehat{\cal Q}_{\mu\,{\cal A}}{}^{\cal B}$ of the $\O(d+1)\times\O(d+1)$ connection ${\cal Q}_{\mu\,{\cal A}}{}^{\cal B}$ of eq.~\eqref{eq:Qd+1} (\textit{i.e.} defining $\widehat{\cal Q}_{\mu\,{\cal A}}{}^{\cal B}$ from the covariantized Maurer-Cartan form ${\cal V}^{-1}D_{\mu}{\cal V}$), we get
 \begin{equation} \label{eq:CSQcov}
  {\rm CS}_{\mu\nu\rho}\left(\widehat{\cal Q}\right) = \frac{2}{3}\,\Omega_{\mu\nu\rho}^{(\rm GS)}-\frac{1}{2}\,e^{2\Phi}\,D_{[\mu}\xi_{M}\left(\H^{-1}\partial_{\nu}\H\right)^{M}{}_{N}D_{\rho]}\xi^{N}- \frac{m}{2}\,e^{2\Phi}\,D_{[\mu}\xi_{M}\F_{\nu\rho]}{}^{M},
 \end{equation}
 of exterior derivative\footnote{Here, we used that $\left[D_{\mu},D_{\nu}\right]\M_{\cal MN} = 4\,F_{\mu\nu}{}^{\cal PQ}\Theta_{\cal PQ\vert (M}{}^{\cal K}\M_{\cal N)K}$ and $F_{\mu\nu}{}^{\cal MN} = 2\,\partial_{[\mu}A_{\nu]}{}^{\cal MN}+4\,A_{[\mu}{}^{\cal PQ}\Theta_{\cal PQ\vert K}{}^{\cal [M}A_{\nu]}{}^{\cal N]K}$.}
 \begin{equation}
    \begin{aligned}
      \partial_{[\mu}{\rm CS}_{\nu\rho\sigma]}\left(\widehat{\cal Q}\right)&=\frac{1}{16}\,\Tr{D_{[\mu}\M D_{\nu}\M^{-1}D_{\rho}\M D_{\sigma]}\M^{-1}\M\eta}\\
      &\quad-2\,\Theta_{\cal MN\vert PQ}\,F_{[\mu\nu}{}^{\cal MN}\left(D_{\rho}\M^{-1} D_{\sigma]}\M\M^{-1}\right)^{\cal PQ}\\
      &\quad+16\,\Theta_{\cal MN\vert PQ}\Theta_{\cal KL\vert R}{}^{\cal P}\,F_{[\mu\nu}{}^{\cal MN}F_{\rho\sigma]}{}^{\cal KL}\M^{\cal RQ}.
    \end{aligned}
  \end{equation}
  Putting this term in the action and eliminating the field-strength using the dualisation relation~\eqref{eq:dual2dermtheta} gives
  \begin{equation}
    \begin{aligned}
    \int\d^{3}x\,m\,\varepsilon^{\mu\nu\rho}\,{\rm CS}_{\mu\nu\rho}\left(\widehat{\cal Q}\right) = \int\d^{3}x\,m\,\bigg[&\varepsilon^{\mu\nu\rho}\,\left(\frac{2}{3}\,\Omega_{\mu\nu\rho}^{(\rm GS)}-\frac{1}{2}\,e^{2\Phi}\,D_{\mu}\xi_{M}\left(\H^{-1}\partial_{\nu}\H\right)^{M}{}_{N}D_{\rho}\xi^{N}\right)\\
    &+\sqrt{-g_{\rm E}}\,m^{2}\,e^{4\Phi}\,D_{\mu}\xi_{M}\H^{MN}D^{\mu}\xi_{N}\bigg].
    \end{aligned}
  \end{equation}
Thus, we can write the action~\eqref{eq:I1secondd+1currents} in terms of the three-form~\eqref{eq:CSQcov}:
\begin{equation}
  \begin{aligned}
    \widetilde{I}''_{1} = \widehat{I}_{1}+&\int\d^{3}x\,\Bigg[\sqrt{-g_{\rm E}}\,\frac{a+b}{8}\,\bigg(m^{2}\Big(4\,e^{2\Phi}\,\partial_{\mu}\Phi\partial^{\mu}\Phi+e^{4\Phi}\,D_{\mu}\xi_{M}\H^{MN}D^{\mu}\xi_{N}\Big)+m^{4}\,e^{6\Phi}\bigg)\\
    &+m\,\varepsilon^{\mu\nu\rho}\bigg(-\frac{a+b}{8}\,{\rm CS}_{\mu\nu\rho}\left(\widehat{\cal Q}\right)+\frac{a-b}{4}\,\Omega_{\mu\nu\rho}^{(\omega_{\rm E})}\bigg)\Bigg].
  \end{aligned}
\end{equation}
We can furthermore express the first line in terms of the $\O(d+1,d+1)$ currents, as done in sec.~\ref{sec:formalcovariance}, and the embedding tensor~\eqref{eq:embeddingm}:
\begin{equation} \label{eq:I1massivedef}
  \begin{aligned}
    \widetilde{I}''_{1} = \widehat{I}_{1}+&\int\d^{3}x\,m\,\Bigg[\sqrt{-g_{\rm E}}\,\frac{a+b}{8}\,\left(\frac{1}{2}\,\theta_{\cal MN}\Big(\widehat{\J}_{\mu}\widehat{\J}^{\mu}\M\Big)^{\cal MN}+\frac{1}{8}\left(\theta_{\cal MN}\M^{\cal MN}\right)^{3}\right)\\
    &+\varepsilon^{\mu\nu\rho}\bigg(-\frac{a+b}{8}\,{\rm CS}_{\mu\nu\rho}\left(\widehat{\cal Q}\right)+\frac{a-b}{4}\,\Omega_{\mu\nu\rho}^{(\omega_{\rm E})}\bigg)\Bigg].
  \end{aligned}
\end{equation}
Thus, the massive deformation following from integrating out the $B$ field induces a gauging of the action~\eqref{eq:I1d+1currentsproj} and additional couplings. Remarkably, these couplings break the $\O(d+1,d+1)$ symmetry only due to the gauging~\eqref{eq:embeddingm} of the shift symmetry; no vector compensator is needed, in contrast to the term $\widehat{I}_{1}$. The first line in eq.~\eqref{eq:I1massivedef} features a deformation of the two-derivative action, and a potential. 
The second line of eq.~\eqref{eq:I1massivedef} is given by Chern-Simons terms based on composite and spin connections. Most interestingly, to leading order in $m$, these Chern-Simons terms are given by
\begin{equation}
  \begin{aligned}
    \int\d^{3}x\,m\,\varepsilon^{\mu\nu\rho}\bigg(-\frac{a+b}{8}\,&{\rm CS}_{\mu\nu\rho}\left(\widehat{\cal Q}\right)+\frac{a-b}{4}\,\Omega_{\mu\nu\rho}^{(\omega_{\rm E})}\bigg) \\
    &= \int\d^{3}x\,m\,\varepsilon^{\mu\nu\rho}\bigg(-\frac{a+b}{8}\,{\rm CS}_{\mu\nu\rho}\left({\cal Q}\right)+\frac{a-b}{4}\,\Omega_{\mu\nu\rho}^{(\omega_{\rm E})}\bigg) + {\cal O}\left(m^{2}\right),
  \end{aligned}
\end{equation}
which is invariant under $\O(d+1,d+1)$. Although the full theory exhibits a breaking of the $\O(d+1,d+1)$ U-duality both by the higher-derivative parameter $\alpha'$ and the mass deformation $m$, for the leading Chern-Simons terms the full $\O(d+1,d+1)$ is restored. The physical meaning of this observation has to be investigated, as well as its extension to more general gaugings.

\section{Conclusions} \label{sec:ccl}

In this paper we have computed the effective action of bosonic and heterotic string theory in three dimensions 
to first order in $\alpha'$ starting from the known four-derivative result in manifestly $\O(d,d)$-invariant form 
and perturbatively dualizing the vector gauge fields into scalars. We have cast the result into a formally $\O(d+1,d+1)$-invariant form upon introduction of a non-dynamical compensator $\veccomp^{\cal M}$. 
The resulting action reveals the intriguing chiral pattern (\ref{chiral}),
showing that in particular the action of the bosonic string can be reconstructed from the heterotic action. 
One may expect that an extension of this structure to higher orders in $\alpha'$ will constrain the potential 
higher order corrections in a similar way.

Another interesting application of the formally $\O(d+1,d+1)$-invariant formulation will be the study of $\O(4,4)$ triality rotations on the resulting action. This would for instance allow one to relate the inequivalent higher order corrections obtained from $T^3$ compactification of chiral ${\cal N}=(2,0)$ and non-chiral ${\cal N}=(1,1)$ supergravity in six dimensions, respectively.
With the latter corresponding to the standard heterotic corrections (see \cite{Chang:2022urm} for a recent discussion) this may be turned into a prediction of the $\alpha'$ corrections of the chiral theory in six dimensions. As this theory arises from ten-dimensional type IIB supergravity compactified on the complex surface K3, this computation would give profitable insights into the higher-derivative corrections in ten dimensions.

The other main result of this paper is the massive deformation identified in sec.~\ref{sec:Bfield} upon integrating out the $B$ field in three dimensions, keeping a constant three-form flux. As exhibited in eq.~(\ref{eq:I1massivedef}) above, this gives rise to a deformation of the target space metric and the scalar potential, as well as to new Chern-Simons terms for composite connections which remarkably exhibit an enhancement to the full $\O(d+1,d+1)$ to first order in the mass parameter. Within the general framework of gauged supergravities, such massive deformations take the form of a particular and somewhat degenerate example of a general gauging. In this respect, our results may be viewed as a glimpse into the structure of the $\alpha'$ corrections of more general gauged supergravities in three dimensions. Of particular interest for holographic applications would be the study of these deformations around the AdS$_3\times S^3$ background. We hope to come back to these issues in the future.

\section*{Acknowledgements}
We thank Guillaume Bossard and Massimo Porrati for comments and discussions. CE is supported by the FWO-Vlaanderen through the project G006119N and by the Vrije Universiteit Brussel through the Strategic Research Program “High-Energy Physics”. OH is supported by the European Research Council (ERC) under the European
Union’s Horizon 2020 research and innovation program (grant agreement No 771862).

\begin{appendix}
\section{\Odd decompositions of \texorpdfstring{$\O(d+1,d+1)$}{O(d+1,d+1)}} \label{app:Od1d1toOdd}

We list in the following the $\Odd$ decomposition of the $\O(d+1,d+1)$ terms used in sec.~\ref{sec:formalcovariance} and of the basis of $\O(d+1,d+1)$-invariant terms carrying four derivatives~\cite{Eloy:2019hnl}.

\begin{equation}
  \Tr{\J_{\mu}\J_{\nu}}=-\Tr{\partial_{\mu}{\cal H}\partial_{\nu}{\cal H}^{-1}}+4\,e^{2\Phi}\,\partial_{\mu}\xi_{M}\H^{MN}\partial_{\nu}\xi_{N}+8\,\partial_{\mu}\Phi\partial_{\nu}\Phi.
\end{equation}
\begin{equation}
  \left(\J_{\mu}\J_{\nu}\M\right)_{++} = e^{4\Phi}\,\partial_{\mu}\xi_{M}\H^{MN}\partial_{\nu}\xi_{N}+4\,e^{2\Phi}\,\partial_{\mu}\Phi\partial_{\nu}\Phi.
\end{equation}

\begin{equation}
  \left(\J_{\mu}\J_{\nu}\eta\right)_{++} = -e^{4\Phi}\,\partial_{\mu}\xi_{M}\partial_{\nu}\xi^{M}.
\end{equation}

\begin{equation}
  \begin{aligned}
    &\Tr{\J_{\mu}\J_{\nu}\J^{\mu}\J^{\nu}} = \Tr{\partial_{\mu}{\cal H}\partial_{\nu}{\cal H}^{-1}\partial^{\mu}{\cal H}\partial^{\nu}{\cal H}^{-1}} -8\,e^{2\Phi}\,\partial_{\mu}\xi_{M}\left(\H^{-1}\partial_{\nu}\H\partial^{\mu}\H^{-1}\right)^{MN}\partial^{\nu}\xi_{N} \\
    &\qquad\quad-16\,e^{2\Phi}\,\partial_{\mu}\Phi\partial_{\nu}\xi_{M}\partial^{\mu}\H^{MN}\partial^{\nu}\xi_{N}+32\,\partial_{\mu}\Phi\partial^{\mu}\Phi\partial_{\nu}\Phi\partial^{\mu}\Phi+4\,e^{4\Phi}\,\partial_{\mu}\xi_{M}\partial_{\nu}\xi^{M}\partial^{\mu}\xi_{N}\partial^{\nu}\xi^{N}\\
    &\qquad\quad +4\,e^{4\Phi}\,\partial_{\mu}\xi_{M}\H^{MN}\partial_{\nu}\xi_{N}\partial^{\mu}\xi_{P}\H^{PQ}\partial^{\nu}\xi^{Q}+32\,e^{2\Phi}\,\partial_{\mu}\Phi\partial_{\nu}\Phi\partial^{\mu}\xi_{M}\H^{MN}\partial^{\nu}\xi_{N}.
  \end{aligned}
\end{equation}
\begin{equation}
  \begin{aligned}
    &\Tr{\J_{\mu}\J^{\mu}\J_{\nu}\J^{\nu}} = \Tr{\partial_{\mu}{\cal H}\partial^{\mu}{\cal H}^{-1}\partial_{\nu}{\cal H}\partial^{\nu}{\cal H}^{-1}} -4\,e^{2\Phi}\,\partial_{\mu}\xi_{M}\left(\H^{-1}\partial_{\nu}\H\partial^{\nu}\H^{-1}\right)^{MN}\partial^{\mu}\xi_{N} \\
    &\qquad\quad-4\,e^{2\Phi}\,\partial_{\mu}\xi_{M}\left(\H^{-1}\partial^{\mu}\H\partial_{\nu}\H^{-1}\right)^{MN}\partial^{\nu}\xi_{N}-16\,e^{2\Phi}\,\partial_{\mu}\Phi\partial^{\mu}\xi_{M}\partial_{\nu}\H^{MN}\partial^{\nu}\xi_{N}\\
    &\qquad\quad+32\,\partial_{\mu}\Phi\partial^{\mu}\Phi\partial_{\nu}\Phi\partial^{\mu}\Phi+2\,e^{4\Phi}\,\partial_{\mu}\xi_{M}\partial^{\mu}\xi^{M}\partial_{\nu}\xi_{N}\partial^{\nu}\xi^{N}+2\,e^{4\Phi}\,\partial_{\mu}\xi_{M}\partial_{\nu}\xi^{M}\partial^{\mu}\xi_{N}\partial^{\nu}\xi^{N}\\
    &\qquad\quad+2\,e^{4\Phi}\,\partial_{\mu}\xi_{M}\H^{MN}\partial^{\mu}\xi_{N}\partial_{\nu}\xi_{P}\H^{PQ}\partial^{\nu}\xi^{Q}+2\,e^{4\Phi}\,\partial_{\mu}\xi_{M}\H^{MN}\partial_{\nu}\xi_{N}\partial^{\mu}\xi_{P}\H^{PQ}\partial^{\nu}\xi^{Q}\\
    &\qquad\quad+16\,e^{2\Phi}\,\partial_{\mu}\Phi\partial^{\mu}\Phi\partial_{\nu}\xi_{M}\H^{MN}\partial^{\nu}\xi_{N}+16\,e^{2\Phi}\,\partial_{\mu}\Phi\partial_{\nu}\Phi\partial^{\mu}\xi_{M}\H^{MN}\partial^{\nu}\xi_{N}.
  \end{aligned}
\end{equation}
\begin{equation}
  \begin{aligned}
    &\Tr{\J_{\mu}\J^{\mu}\J_{\nu}\J^{\nu}{\cal M}\eta} = \Tr{\partial_{\mu}{\cal H}\partial^{\mu}{\cal H}^{-1}\partial_{\nu}{\cal H}\partial^{\nu}{\cal H}^{-1}{\cal H}\eta} -4\,e^{2\Phi}\,\partial_{\mu}\xi_{M}\left(\partial_{\nu}\H^{-1}\partial^{\nu}\H\right)^{M}{}_{N}\partial^{\mu}\xi^{N}  \\
    &\qquad\quad + 4\,e^{2\Phi}\,\partial_{\mu}\xi_{M}\left(\partial^{\mu}\H^{-1}\partial_{\nu}\H\right)^{M}{}_{N}\partial^{\nu}\xi^{N}-16\,e^{2\Phi}\,\partial_{\mu}\Phi\partial^{\mu}\xi_{M}\left(\H^{-1}\partial_{\nu}\H\right)^{M}{}_{N}\partial^{\nu}\xi^{N}\\
    &\qquad\quad+4\,e^{4\Phi}\,\partial_{\mu}\xi_{M}\H^{MN}\partial_{\nu}\xi_{N}\partial^{\mu}\xi_{P}\partial^{\nu}\xi^{P}-4\,e^{4\Phi}\,\partial_{\mu}\xi_{M}\H^{MN}\partial^{\mu}\xi_{N}\partial_{\nu}\xi_{P}\partial^{\nu}\xi^{P} \\
    &\qquad\quad-16\,e^{2\Phi}\,\partial_{\mu}\Phi\partial^{\mu}\Phi\partial_{\nu}\xi_{M}\partial^{\nu}\xi^{M}+16\,e^{2\Phi}\,\partial_{\mu}\Phi\partial_{\nu}\Phi\partial^{\mu}\xi_{M}\partial^{\nu}\xi^{M}.
  \end{aligned}
\end{equation}

\begin{equation}
  \begin{aligned}
    \left(\J_{\mu}\J_{\nu}\J^{\mu}\J^{\nu}\M\right)_{++} &= -\,e^{4\Phi}\,\partial_{\mu}\xi_{M}\left(\H^{-1}\partial_{\nu}\H\partial^{\mu}\H^{-1}\right)^{MN}\partial^{\nu}\xi_{N}-4\,e^{4\Phi}\,\partial_{\mu}\Phi \,\partial_{\nu}\xi_{M}\partial^{\mu}\H^{MN}\partial_{\nu}\xi_{N} \\
    &\quad+16\,e^{2\Phi}\,\partial_{\mu}\Phi\partial^{\mu}\Phi\partial_{\nu}\Phi\partial^{\nu}\Phi + e^{6\,\Phi}\,\partial_{\mu}\xi_{M}\H^{MN}\partial_{\nu}\xi_{N}\,\partial^{\mu}\xi_{P}\H^{PQ}\partial^{\nu}\xi_{Q}\\
    &\quad+e^{6\Phi}\,\partial_{\mu}\xi_{M}\partial_{\nu}\xi^{M}\,\partial^{\mu}\xi_{N}\partial^{\nu}\xi^{N}+ 12\,e^{4\Phi}\,\partial_{\mu}\Phi\partial_{\nu}\Phi \partial^{\mu}\xi_{M}\H^{MN}\partial^{\nu}\xi_{N}.
  \end{aligned}
\end{equation}

\begin{equation}
  \begin{aligned}
    \left(\J_{\mu}\J_{\nu}\J^{\mu}\J^{\nu}\eta\right)_{++} &= e^{4\Phi}\,\partial_{\mu}\xi_{M}\left(\partial_{\nu}\H^{-1}\partial^{\mu}\H\right)^{M}{}_{N}\partial^{\nu}\xi^{N}- 2\,e^{6\Phi}\,\partial_{\mu}\xi_{M}\partial_{\nu}\xi^{M}\,\partial^{\mu}\xi_{P}\H^{PQ}\partial^{\nu}\xi_{Q} \\
    &\quad -4\,e^{4\Phi}\,\partial_{\mu}\Phi\partial_{\nu}\Phi \partial^{\mu}\xi_{M}\partial^{\nu}\xi^{M}.
  \end{aligned}
\end{equation}

\begin{equation}
  \begin{aligned}
    \left(\J_{\mu}\J^{\mu}\J_{\nu}\J^{\nu}\M\right)_{++} &=-\,e^{4\Phi}\,\partial_{\mu}\xi_{M}\left(\H^{-1}\partial^{\mu}\H\partial^{\nu}\H^{-1}\right)^{MN}\partial_{\nu}\xi_{N}- 4\,e^{4\Phi}\,\partial_{\mu}\Phi \,\partial^{\mu}\xi_{M}\partial_{\nu}\H^{MN}\partial^{\nu}\xi_{N} \\
    &\quad+ 16\,e^{2\Phi}\,\partial_{\mu}\Phi\partial^{\mu}\Phi\partial_{\nu}\Phi\partial^{\nu}\Phi + e^{6\Phi}\,\partial_{\mu}\xi_{M}\H^{MN}\partial^{\mu}\xi_{N}\,\partial_{\nu}\xi_{P}\H^{PQ}\partial^{\nu}\xi_{Q}  \\
    &\quad +e^{6\Phi}\,\partial_{\mu}\xi_{M}\partial^{\mu}\xi^{M}\,\partial_{\nu}\xi_{N}\partial^{\nu}\xi^{N} + 4\,e^{4\Phi}\,\partial_{\mu}\Phi\partial_{\nu}\Phi\, \partial^{\mu}\xi_{M}\H^{MN}\partial^{\nu}\xi_{N}  \\
    &\quad + 8\,e^{4\Phi}\,\partial_{\mu}\Phi\partial^{\mu}\Phi \, \partial_{\nu}\xi_{M}\H^{MN}\partial^{\nu}\xi_{N}.
  \end{aligned}
\end{equation}

\begin{equation}
  \begin{aligned}
    \left(\J_{\mu}\J^{\mu}\J_{\nu}\J^{\nu}\eta\right)_{++} &= e^{4\Phi}\,\partial_{\mu}\xi_{M}\left(\partial^{\mu}\H^{-1}\partial_{\nu}\H\right)^{M}{}_{N}\partial^{\nu}\xi^{N} +4\,e^{4\Phi}\,\partial_{\mu}\Phi \, \partial^{\mu}\xi_{M}\left(\partial_{\nu}\H^{-1}\H\right)^{M}{}_{N}\partial^{\nu}\xi^{N} \\
    &\quad- 2\,e^{6\Phi}\,\partial_{\mu}\xi_{M}\partial^{\mu}\xi^{M}\,\partial_{\nu}\xi_{P}\H^{PQ}\partial^{\nu}\xi_{Q} +4\,e^{4\Phi}\, \partial_{\mu}\Phi\partial_{\nu}\Phi \partial^{\mu}\xi_{M}\partial^{\nu}\xi^{M}\\
    &\quad-8\,e^{4\Phi}\, \partial_{\mu}\Phi\partial^{\mu}\Phi \partial_{\nu}\xi_{M}\partial^{\nu}\xi^{M}.
  \end{aligned}
\end{equation}

\begin{equation}
  \begin{aligned}
    \left(\J_{\mu}\J_{\nu}\J^{\nu}\J^{\mu}\M\right)_{++} &= -\,e^{4\Phi}\,\partial_{\mu}\xi_{M}\left(\H^{-1}\partial_{\nu}\H\partial^{\nu}\H^{-1}\right)^{MN}\partial^{\mu}\xi_{N} - 4\,e^{4\Phi}\,\partial_{\mu}\Phi \,\partial^{\mu}\xi_{M}\partial_{\nu}\H^{MN}\partial^{\nu}\xi_{N} \\
    &\quad+16\,e^{2\Phi}\,\partial_{\mu}\Phi\partial^{\mu}\Phi\partial_{\nu}\Phi\partial^{\nu}\Phi + e^{6\Phi}\,\partial_{\mu}\xi_{M}\H^{MN}\partial_{\nu}\xi_{N}\,\partial^{\mu}\xi_{P}\H^{PQ}\partial^{\nu}\xi_{Q}  \\
    &\quad+e^{6\Phi}\,\partial_{\mu}\xi_{M}\partial_{\nu}\xi^{M}\,\partial^{\mu}\xi_{N}\partial^{\nu}\xi^{N} + 8\,e^{4\Phi}\,\partial_{\mu}\Phi\partial_{\nu}\Phi\, \partial^{\mu}\xi_{M}\H^{MN}\partial^{\nu}\xi_{N} \\
    &\quad +4\,e^{4\Phi}\,\partial_{\mu}\Phi\partial^{\mu}\Phi \, \partial_{\nu}\xi_{M}\H^{MN}\partial^{\nu}\xi_{N}.
  \end{aligned}
\end{equation}

\begin{equation}
  \begin{aligned}
    \left(\J_{\mu}\J_{\nu}\J^{\nu}\J^{\mu}\eta\right)_{++} &= e^{4\Phi}\,\partial_{\mu}\xi_{M}\left(\partial_{\nu}\H^{-1}\partial^{\nu}\H\right)^{M}{}_{N}\partial^{\mu}\xi^{N}-4\,e^{4\Phi}\,\partial_{\mu}\Phi \, \partial^{\mu}\xi_{M}\left(\partial_{\nu}\H^{-1}\H\right)^{M}{}_{N}\partial^{\nu}\xi^{N}   \\
    &\quad- 2\,e^{6\Phi}\,\partial_{\mu}\xi_{M}\partial_{\nu}\xi^{M}\,\partial^{\mu}\xi_{P}\H^{PQ}\partial^{\nu}\xi_{Q} -8\,e^{4\Phi}\, \partial_{\mu}\Phi\partial_{\nu}\Phi \partial^{\mu}\xi_{M}\partial^{\nu}\xi^{M}\\
    &\quad+4\,e^{4\Phi}\, \partial_{\mu}\Phi\partial^{\mu}\Phi \partial_{\nu}\xi_{M}\partial^{\nu}\xi^{M}.
  \end{aligned}
\end{equation}

\section{Four-derivative action with mass deformation} \label{app:4derBfield}
We detail here the computation of the action~\eqref{eq:I1secondd+1currents}. We first give the rules for field redefinitions as modified by the mass deformation of the two-derivative action. We then move to Einstein frame, use field redefinitions to convert all second order derivatives into product of first order derivatives and finally dualise the vector fields.

\subsection{Field redefinitions}
The two-derivative action in Einstein frame with $B$ field integrated out is given in eq.~\eqref{eq:secondI0einstein}. Its equations of motion are
\begin{equation}
    \delta I''_{0} = \alpha'\,\int \d^{3}x\,\sqrt{-g_{\rm E}}\bigg[\delta\Phi\,E_{\Phi} + \delta g^{\mu\nu}_{\rm E} E_{g\,\mu\nu}+ \Tr{\delta \H^{-1}\,E_{\cal H}}+\delta\A_{\mu}{}^{M}E_{\A}{}^{\mu}{}_{M}\bigg],
  \end{equation}
  where
  \begin{subnumcases}{\label{eq:eomBfield}}
    E_{\Phi} = 2\,\Box\Phi +\frac{1}{2} e^{-2\Phi}\,\F_{\mu\nu}{}^{M}{\cal H}_{MN}\F^{\mu\nu\,N} -2\,m^{2}\,e^{4\Phi}, \label{eq:eomphiBfield} \\
    E_{g\,\mu\nu} = R_{{\rm E}\,\mu\nu}-\partial_{\mu}\Phi\partial_{\nu}\Phi + \frac{1}{8}\,\Tr{\partial_{\mu}\H\,\partial_{\nu}\H^{-1}}-\frac{1}{2}\,e^{-2\Phi}\,\F_{\mu\rho}{}^{M}{\cal H}_{MN}\F_{\nu}{}^{\rho\,N} \nonumber  \\ 
    \qquad \quad - \frac{1}{2}g_{{\rm E}\, \mu\nu} \left(R_{\rm E}-\partial_{\rho}\Phi\,\partial^{\rho}\Phi+\frac{1}{8}\,\Tr{\partial_{\rho}\H\,\partial^{\rho}\H^{-1}}-\frac{1}{4}\,e^{-2\Phi}\,\F_{\rho\sigma}{}^{M}{\cal H}_{MN}\F^{\rho\sigma\,N} -\frac{1}{2}\,m^{2}\,e^{4\Phi}\right), \label{eq:eomgBfield} \\
    E_{{\cal H}\,MN}  = -\frac{1}{4}\,\bigg[\Box\,\H_{MN}+\left(\H\partial_{\mu}\H^{-1}\,\partial^{\mu}\H\right)_{MN} + \frac{1}{2}\,e^{-2\Phi}\,\F_{\mu\nu\,M}\F^{\mu\nu}{}_{N} - \frac{1}{2}\,e^{-2\Phi}\,\F_{\mu\nu}{}^{P}\H_{PM}\F^{\mu\nu\,Q}\H_{QN}\bigg], \label{eq:eomSBfield} \\
    E_{\A}{}^{\mu}{}_{M} = e^{-2\Phi}\,\nabla_{\nu}\F^{\nu\mu\,N}\H_{NM}-2\,e^{-2\Phi}\,\nabla_{\nu}\Phi\,\F^{\nu\mu\,N}\H_{NM}+e^{-2\Phi}\,\F^{\nu\mu\,N}\nabla_{\nu}\H_{NM}+\frac{m}{2}\,\epsilon^{\mu\nu\rho}\F_{\nu\rho\,M} . \label{eq:eomABfield}
  \end{subnumcases}
  The resulting field redefinitions are as follows:
  \begin{equation}  \label{eq:fieldredefBfield}
    \begin{aligned}
    \Box\Phi&\longrightarrow Q_{\Phi} = -\frac{1}{4} e^{-2\Phi}\,\F_{\mu\nu}{}^{M}{\cal H}_{MN}\F^{\mu\nu\,N} +m^{2}\,e^{4\Phi}, \\
    R_{{\rm E}\,\mu\nu}&\longrightarrow Q_{g\,\mu\nu} = \partial_{\mu}\Phi\partial_{\nu}\Phi - \frac{1}{8}\,\Tr{\partial_{\mu}\H\,\partial_{\nu}\H^{-1}} +\frac{1}{2}\,e^{-2\Phi}\,\F_{\mu\rho}{}^{M}{\cal H}_{MN}\F_{\nu}{}^{\rho\,N}\\
    &\qquad\quad\qquad-\frac{1}{4}\,g_{{\rm E}\,\mu\nu}\,e^{-2\Phi}\,\F_{\rho\sigma}{}^{M}{\cal H}_{MN}\F^{\rho\sigma\,N}+\frac{1}{2}\,g_{{\rm E}\,\mu\nu}\,m^{2}\,e^{4\Phi}, \\
    \Box\,\H_{MN}&\longrightarrow Q_{\H\,MN} = -\left(\H\,\partial_{\mu}\H^{-1}\partial^{\mu}\H\right)_{MN} - \frac{1}{2}\,e^{-2\Phi}\,\F_{\mu\nu\,M}\F^{\mu\nu}{}_{N}\\
    &\qquad\qquad\qquad + \frac{1}{2}\,e^{-2\Phi}\,\F_{\mu\nu}{}^{P}\H_{PM}\F^{\mu\nu\,Q}\H_{QN}, \\
    \nabla_{\nu}\F^{\nu\mu\,M}&\longrightarrow Q_{\A}{}^{\mu M} =  2\,\partial_{\nu}\Phi\,\F^{\nu\mu\,M}-\F^{\nu\mu\,N}\left(\partial_{\nu}{\cal H}{\cal H}^{-1}\right)_{N}{}^{M}-\frac{m}{2}\,e^{2\Phi}\,\epsilon^{\mu\nu\rho}\,\F_{\nu\rho\,N}\H^{NM}.\\
    \end{aligned}
  \end{equation}
  \subsection{Einstein frame}
  We write the action~\eqref{eq:I1second} in Einstein frame ($g_{\mu\nu}\rightarrow g_{{\rm E}\,\mu\nu} = e^{-2\Phi}g_{\mu\nu}$):
  \begin{equation} \label{eq:I1secondEinstein}
    \begin{aligned}
      I''_{1}=&\int\d^{3}x\,\sqrt{-g_{\rm E}}\Bigg[ \\
      &-\frac{a+b}{8}\,e^{-2\Phi}\,\bigg(4\,R_{{\rm E}\mu\nu}R_{\rm E}{}^{\mu\nu} - R_{\rm E}^{2} - 8\,R_{{\rm E}\,\mu\nu}\nabla^{\mu}\nabla^{\nu}\Phi+8\,R_{{\rm E}\,\mu\nu}\,\partial^{\mu}\Phi\partial^{\nu}\Phi -4\,R_{\rm E}\,\partial_{\mu}\Phi\partial^{\mu}\Phi\\
      &\qquad+4\,\nabla_{\mu}\nabla_{\nu}\Phi\nabla^{\mu}\nabla^{\nu}\Phi +4\,\Box\Phi\Box\Phi-8\,\nabla_{\mu}\nabla_{\nu}\Phi \partial^{\mu}\Phi\partial^{\nu}\Phi +8\,\Box\Phi\partial_{\mu}\Phi\partial^{\mu}\Phi+4\,\partial_{\mu}\Phi\partial^{\mu}\Phi\partial_{\nu}\Phi\partial^{\nu}\Phi \\
      &\qquad+\frac{1}{16}\,\Tr{\partial_{\mu}\H\partial_{\nu}\H^{-1}\partial^{\mu}\H\partial^{\nu}\H^{-1}} - \frac{1}{32}\, \Tr{\partial_{\mu}\H\partial_{\nu}\H^{-1}} \Tr{\partial^{\mu}\H\partial^{\nu}\H^{-1}}  \\
      &\qquad+\frac{1}{8}\,e^{-4\Phi}\,{\F_{\mu\nu}}^{M}{\F_{\rho\sigma}}_{M}\F^{\mu\rho\,N}{\F^{\nu\sigma}}_{N}+\frac{1}{8}\,e^{-4\,\Phi}\, {\F_{\mu\nu}}^{M}\H_{MN}\F_{\rho\sigma}{}^{N}\F^{\mu\rho\,P}\H_{PQ}\F^{\nu\sigma\,Q}\\
      &\qquad -\frac{1}{2}\,e^{-4\Phi}\,{\F_{\mu\nu}}^{M}\H_{MN}\F^{\mu\rho\,N}\F^{\nu\sigma\,P}\H_{PQ}\F_{\rho\sigma}{}^{Q}+\frac{1}{2}\,e^{-2\Phi}\left(R_{\rm E}+2\,\partial_{\mu}\Phi\partial^{\mu}\Phi\right)\F_{\rho\sigma}{}^{M}\H_{MN}\F^{\rho\sigma\,N} \\
      &\qquad-2\,e^{-2\Phi}\left(R_{{\rm E}\,\mu\nu}-\nabla_{\mu}\nabla_{\nu}\Phi+\partial_{\mu}\Phi\partial_{\nu}\Phi\right)\F^{\mu\rho\,M}\H_{MN}\F^{\nu}{}_{\rho}{}^{N}\\
      &\qquad-\frac{1}{2}\,e^{-2\Phi}\,{\F_{\mu\nu}}^{M}\left(\H\partial_{\rho}\H^{-1}\partial^{\nu}\H\right)_{MN} \F^{\mu\rho\,N} + \frac{1}{4}\,e^{-2\Phi}\,\F^{\mu\rho\,M}\H_{MN}\F^{\nu}{}_{\rho}{}^{N}\,\Tr{\partial_{\mu}\H\partial_{\nu}\H^{-1}}\\
      &\qquad+m\,e^{2\Phi}\,\epsilon^{\mu\nu\rho}\left(\frac{2}{3}\,\Omega_{\mu\nu\rho}^{\rm (GS)}-\frac{1}{2}\,e^{-2\Phi}\,\F_{\mu\sigma}{}^{M}\left(\H\partial_{\nu}\H^{-1}\right)_{M}{}^{N}\F_{\rho}{}^{\sigma}{}_{N}\right)-\frac{5}{4}\,m^{4}\,e^{8\Phi}\\
      &\qquad+m^{2}\,e^{4\Phi}\left(R_{\rm E}-4\,\Box\Phi-2\,\partial_{\mu}\Phi\partial^{\mu}\Phi-\frac{1}{4}\,\Tr{\partial_{\mu}\H\partial^{\mu}\H^{-1}}+\frac{3}{4}\,e^{-2\Phi}\,\F_{\mu\nu}{}^{M}\H_{MN}\F^{\mu\nu\,N}\right)\bigg)\\
      &+\frac{a-b}{4}\,e^{-2\Phi}\bigg(-\frac{1}{16}\,\Tr{\partial_{\mu}\H\partial^{\mu}\H^{-1}\partial_{\nu}\H\partial^{\nu}\H^{-1}\H\eta} - \frac{1}{16}\,e^{-4\Phi}\,\F_{\mu\nu}{}^{M}\F_{\rho\sigma\,M}\,\F^{\mu\nu\,P}\H_{PQ}\F^{\rho\sigma\,Q}\\
      &\qquad-\frac{1}{4}\,e^{-2\Phi}\,R_{\rm E}\,\F_{\mu\nu}{}^{M}\F^{\mu\nu}{}_{M} +e^{-2\Phi}\,R_{\rm E\,\mu\nu}\,\F^{\mu\rho\,M}\F^{\nu}{}_{\rho\,M}-e^{-2\Phi}\,\nabla_{\mu}\nabla_{\nu}\Phi\,\F^{\mu\rho\,M}\F^{\nu}{}_{\rho\,M}\\
      &\qquad-\frac{1}{2}\,e^{-2\Phi}\,\partial_{\rho}\Phi\partial^{\rho}\Phi\,\F_{\mu\nu}{}^{M}\F^{\mu\nu}{}_{M}+e^{-2\Phi}\,\partial_{\mu}\Phi\partial_{\nu}\Phi\,\F^{\mu\rho\,M}\F^{\nu}{}_{\rho\,M}\\ 
      &\qquad+\frac{1}{8}\,e^{-2\Phi}\,\F_{\mu\nu}{}^{M}\left(\partial_{\rho}\H\partial^{\rho}\H^{-1}\right)_{M}{}^{N}\F^{\mu\nu}{}_{N}+\frac{1}{4}\,e^{-2\Phi}\,\F_{\mu\nu}{}^{M}\left(\partial^{\mu}\H\partial_{\rho}\H^{-1}\right)_{M}{}^{N}\F^{\nu\rho}{}_{N}\\
      &\qquad+m\,e^{2\Phi}\,\epsilon^{\mu\nu\rho}\left(\Omega_{\mu\nu\rho}^{(\omega_{\rm E})}+\frac{1}{4}\,e^{-2\Phi}\,\F_{\mu\sigma}{}^{M}\partial^{\sigma}\H_{MN}\F_{\nu\rho}{}^{N}\right)+\frac{m^{2}}{8}\,e^{2\Phi}\,\F_{\mu\nu}{}^{M}\F^{\mu\nu}{}_{M}\bigg)\Bigg].
    \end{aligned}
  \end{equation}

  \subsection{Integrations by part and field redefinitions}
  We now convert all second order derivatives in eq.~\eqref{eq:I1secondEinstein} into products of first order derivatives: we first use integrations by part so that all the second order derivatives appear in the leading two-derivative contribution of the equations of motion~\eqref{eq:eomBfield}, and obtain\footnote{Note that here, contrary to what we did in sec.~\ref{sec:I1}, we use partial integrations and field redefinitions before dualising the vector fields.}
  \begin{equation}
    \begin{aligned}
      I''_{1}=&\int\d^{3}x\,\sqrt{-g_{\rm E}}\Bigg[ \\
      &-\frac{a+b}{8}\,e^{-2\Phi}\,\bigg(4\,\partial_{\mu}\Phi\partial^{\mu}\Phi\partial_{\nu}\Phi\partial^{\nu}\Phi+\frac{1}{16}\,\Tr{\partial_{\mu}\H\partial_{\nu}\H^{-1}\partial^{\mu}\H\partial^{\nu}\H^{-1}}  \\
      &\qquad - \frac{1}{32}\, \Tr{\partial_{\mu}\H\partial_{\nu}\H^{-1}} \Tr{\partial^{\mu}\H\partial^{\nu}\H^{-1}} +\frac{1}{8}\,e^{-4\,\Phi}\, {\F_{\mu\nu}}^{M}\H_{MN}\F_{\rho\sigma}{}^{N}\F^{\mu\rho\,P}\H_{PQ}\F^{\nu\sigma\,Q}\\
      &\qquad+\frac{1}{8}\,e^{-4\Phi}\,{\F_{\mu\nu}}^{M}{\F_{\rho\sigma}}_{M}\F^{\mu\rho\,N}{\F^{\nu\sigma}}_{N}-\frac{1}{2}\,e^{-4\Phi}\,{\F_{\mu\nu}}^{M}\H_{MN}\F^{\mu\rho\,N}\F^{\nu\sigma\,P}\H_{PQ}\F_{\rho\sigma}{}^{Q} \\
      &\qquad-e^{-2\Phi}\,\partial_{\mu}\Phi\partial^{\mu}\Phi\,\F_{\rho\sigma}{}^{M}\H_{MN}\F^{\rho\sigma\,N}+6\,e^{-2\Phi}\,\partial_{\mu}\Phi\partial_{\nu}\Phi\,\F^{\mu\rho\,M}\H_{MN}\F^{\nu}{}_{\rho}{}^{N}\\
      &\qquad-\frac{1}{2}\,e^{-2\Phi}\,{\F_{\mu\nu}}^{M}\left(\H\partial_{\rho}\H^{-1}\partial^{\nu}\H\right)_{MN} \F^{\mu\rho\,N} + \frac{1}{4}\,e^{-2\Phi}\,\F^{\mu\rho\,M}\H_{MN}\F^{\nu}{}_{\rho}{}^{N}\,\Tr{\partial_{\mu}\H\partial_{\nu}\H^{-1}}\\
      &\qquad-2\,e^{-2\Phi}\,\partial_{\nu}\Phi\,\F^{\mu\rho\,M}\partial_{\mu}\H_{MN}\F^{\nu}{}_{\rho}{}^{N} +\frac{1}{2}\,e^{-2\Phi}\,\partial_{\rho}\Phi\,\F_{\mu\nu}{}^{M}\partial^{\rho}\H_{MN}\F^{\mu\nu\,N} \\
      &\qquad+m\,e^{2\Phi}\,\epsilon^{\mu\nu\rho}\left(\frac{2}{3}\,\Omega_{\mu\nu\rho}^{\rm (GS)}-\frac{1}{2}\,e^{-2\Phi}\,\F_{\mu\sigma}{}^{M}\left(\H\partial_{\nu}\H^{-1}\right)_{M}{}^{N}\F_{\rho}{}^{\sigma}{}_{N}\right)-\frac{5}{4}\,m^{4}\,e^{8\Phi}\\
      &\qquad+m^{2}\,e^{4\Phi}\left(-2\,\partial_{\mu}\Phi\partial^{\mu}\Phi-\frac{1}{4}\,\Tr{\partial_{\mu}\H\partial^{\mu}\H^{-1}}+\frac{3}{4}\,e^{-2\Phi}\,\F_{\mu\nu}{}^{M}\H_{MN}\F^{\mu\nu\,N}\right)\\
      &\qquad +4\,R_{{\rm E}\mu\nu}R_{\rm E}{}^{\mu\nu} - R_{\rm E}^{2}-12\,R_{{\rm E}\,\mu\nu}\,\partial^{\mu}\Phi\partial^{\nu}\Phi +4\,R_{\rm E}\,\partial_{\mu}\Phi\partial^{\mu}\Phi-4\,R_{\rm E}\,\Box\Phi+8\,\Box\Phi\Box\Phi\\
      &\qquad+\frac{1}{2}\,e^{-2\Phi}\,R_{\rm E}\,\F_{\mu\nu}{}^{M}\H_{MN}\F^{\mu\nu\,N}-2\,e^{-2\Phi}\,R_{{\rm E}\,\mu\nu}\,\F^{\mu\rho\,M}\H_{MN}\F^{\nu}{}_{\rho}{}^{N}+m^{2}\,e^{4\Phi}\left(R_{\rm E}-4\,\Box\Phi\right)\\
      &\qquad-2\,e^{-2\Phi}\,\partial_{\nu}\Phi\,\nabla_{\mu}\F^{\mu\rho\,M}\H_{MN}\F^{\nu}{}_{\rho}{}^{N}+\frac{1}{2}\,e^{-2\Phi}\,\Box\Phi\,\F_{\mu\nu}{}^{M}\H_{MN}\F^{\mu\nu\,N}\bigg)\\
      &+\frac{a-b}{4}\,e^{-2\Phi}\bigg(-\frac{1}{16}\,\Tr{\partial_{\mu}\H\partial^{\mu}\H^{-1}\partial_{\nu}\H\partial^{\nu}\H^{-1}\H\eta} - \frac{1}{16}\,e^{-4\Phi}\,\F_{\mu\nu}{}^{M}\F_{\rho\sigma\,M}\,\F^{\mu\nu\,P}\H_{PQ}\F^{\rho\sigma\,Q}\\
      &\qquad-3\,e^{-2\Phi}\,\partial_{\mu}\Phi\partial_{\nu}\Phi\,\F^{\mu\rho\,M}\F^{\nu}{}_{\rho\,M}+\frac{1}{2}\,e^{-2\Phi}\,\partial_{\rho}\Phi\partial^{\rho}\Phi\,\F_{\mu\nu}{}^{M}\F^{\mu\nu}{}_{M}\\ 
      &\qquad+\frac{1}{8}\,e^{-2\Phi}\,\F_{\mu\nu}{}^{M}\left(\partial_{\rho}\H\partial^{\rho}\H^{-1}\right)_{M}{}^{N}\F^{\mu\nu}{}_{N}+\frac{1}{4}\,e^{-2\Phi}\,\F_{\mu\nu}{}^{M}\left(\partial^{\mu}\H\partial_{\rho}\H^{-1}\right)_{M}{}^{N}\F^{\nu\rho}{}_{N}\\
      &\qquad+m\,e^{2\Phi}\,\epsilon^{\mu\nu\rho}\,\left(\Omega_{\mu\nu\rho}^{(\omega_{\rm E})}+\frac{1}{4}\,e^{-2\Phi}\,\F_{\mu\sigma}{}^{M}\partial^{\sigma}\H_{MN}\F_{\nu\rho}{}^{N}\right)+\frac{m^{2}}{8}\,e^{2\Phi}\,\F_{\mu\nu}{}^{M}\F^{\mu\nu}{}_{M}\\
      &\qquad-\frac{1}{4}\,e^{-2\Phi}\,R_{\rm E}\,\F_{\mu\nu}{}^{M}\F^{\mu\nu}{}_{M} +e^{-2\Phi}\,R_{\rm E\,\mu\nu}\,\F^{\mu\rho\,M}\F^{\nu}{}_{\rho\,M}-\frac{1}{4}\,e^{-2\Phi}\,\Box\Phi\,\F_{\mu\nu}{}^{M}\F^{\mu\nu}{}_{M}\\
      &\qquad+e^{-2\Phi}\,\nabla_{\mu}\F^{\mu\rho}\,\partial_{\nu}\Phi\,\F^{\nu}{}_{\rho\,M}\bigg)\Bigg].
    \end{aligned}
  \end{equation}
  Using the field redefinitions~\eqref{eq:fieldredefBfield}, we get
  \begin{equation}
    \begin{aligned}
      I''_{1}=&\int\d^{3}x\,\sqrt{-g_{\rm E}}\Bigg[ \\
      &-\frac{a+b}{8}\,e^{-2\Phi}\,\bigg(\frac{1}{16}\,\Tr{\partial_{\mu}\H\partial_{\nu}\H^{-1}\partial^{\mu}\H\partial^{\nu}\H^{-1}} +\frac{1}{32}\, \Tr{\partial_{\mu}\H\partial_{\nu}\H^{-1}} \Tr{\partial^{\mu}\H\partial^{\nu}\H^{-1}}  \\
      &\qquad -\frac{1}{64}\, \Tr{\partial_{\mu}\H\partial^{\mu}\H^{-1}} \Tr{\partial_{\nu}\H\partial^{\nu}\H^{-1}} +\frac{1}{8}\,e^{-4\,\Phi}\, {\F_{\mu\nu}}^{M}\H_{MN}\F_{\rho\sigma}{}^{N}\F^{\mu\rho\,P}\H_{PQ}\F^{\nu\sigma\,Q}\\
      &\qquad+\frac{1}{8}\,e^{-4\Phi}\,{\F_{\mu\nu}}^{M}{\F_{\rho\sigma}}_{M}\F^{\mu\rho\,N}{\F^{\nu\sigma}}_{N}-\frac{1}{2}\,e^{-4\Phi}\,{\F_{\mu\nu}}^{M}\H_{MN}\F^{\mu\rho\,N}\F^{\nu\sigma\,P}\H_{PQ}\F_{\rho\sigma}{}^{Q} \\
      &\qquad+\frac{3}{16}\,e^{-4\Phi}\,\F_{\mu\nu}{}^{M}\H_{MN}\F^{\mu\nu\,N}\F_{\rho\sigma}{}^{P}\H_{PQ}\F^{\rho\sigma\,Q}-\partial_{\mu}\Phi\partial^{\mu}\Phi\partial_{\nu}\Phi\partial^{\nu}\Phi\\
      &\qquad +\frac{1}{2}\,\partial_{\mu}\Phi\partial_{\nu}\Phi\,\Tr{\partial^{\mu}\H\partial^{\nu}\H^{-1}}-\frac{1}{4}\,\partial_{\mu}\Phi\partial^{\mu}\Phi\,\Tr{\partial_{\nu}\H\partial^{\nu}\H^{-1}} \\
      &\qquad+e^{-2\Phi}\,\partial_{\mu}\Phi\partial^{\mu}\Phi\,\F_{\rho\sigma}{}^{M}\H_{MN}\F^{\rho\sigma\,N}-2\,e^{-2\Phi}\,\partial_{\mu}\Phi\partial_{\nu}\Phi\,\F^{\mu\rho\,M}\H_{MN}\F^{\nu}{}_{\rho}{}^{N}\\
      &\qquad-\frac{1}{2}\,e^{-2\Phi}\,{\F_{\mu\nu}}^{M}\left(\H\partial_{\rho}\H^{-1}\partial^{\nu}\H\right)_{MN} \F^{\mu\rho\,N}  +\frac{1}{2}\,e^{-2\Phi}\,\partial_{\rho}\Phi\,\F_{\mu\nu}{}^{M}\partial^{\rho}\H_{MN}\F^{\mu\nu\,N} \\
      &\qquad+m\,e^{2\Phi}\,\epsilon^{\mu\nu\rho}\left(\frac{2}{3}\,\Omega_{\mu\nu\rho}^{\rm (GS)}-\frac{1}{2}\,e^{-2\Phi}\,\F_{\mu\sigma}{}^{M}\left(\H\partial_{\nu}\H^{-1}\right)_{M}{}^{N}\F_{\rho}{}^{\sigma}{}_{N}+e^{-2\Phi}\,\partial_{\sigma}\Phi\,\F_{\mu\nu}{}^{M}\F^{\sigma}{}_{\rho\,M}\right)\\
      &\qquad-4\,m^{2}\,e^{4\Phi}\,\partial_{\mu}\Phi\partial^{\mu}\Phi-m^{4}\,e^{8\Phi}\bigg)\\
      &+\frac{a-b}{4}\,e^{-2\Phi}\bigg(-\frac{1}{16}\,\Tr{\partial_{\mu}\H\partial^{\mu}\H^{-1}\partial_{\nu}\H\partial^{\nu}\H^{-1}\H\eta} - \frac{1}{16}\,e^{-4\Phi}\,\F_{\mu\nu}{}^{M}\F_{\rho\sigma\,M}\,\F^{\mu\nu\,P}\H_{PQ}\F^{\rho\sigma\,Q}\\
      &\qquad-\frac{1}{8}\,e^{-4\Phi}\,\F_{\mu\nu}{}^{M}\F^{\mu\nu}{}_{M}\,\F_{\rho\sigma}{}^{P}\H_{PQ}\F^{\rho\sigma\,Q}+\frac{1}{2}\,e^{-4\Phi}\,\F_{\mu\rho}{}^{M}\F_{\nu}{}^{\rho}{}_{M}\,\F^{\mu\sigma\,P}\H_{PQ}\F^{\nu}{}_{\sigma}{}^{Q} \\
      &\qquad+\frac{1}{4}\,e^{-2\Phi}\,\partial_{\rho}\Phi\partial^{\rho}\Phi\,\F_{\mu\nu}{}^{M}\F^{\mu\nu}{}_{M}+\frac{1}{32}\,e^{-2\Phi}\,\Tr{\partial_{\rho}\H\partial^{\rho}\H^{-1}}\,\F_{\mu\nu}{}^{M}\F^{\mu\nu}{}_{M}\\
      &\qquad -\frac{1}{8}\,e^{-2\Phi}\,\Tr{\partial_{\mu}\H\partial_{\nu}\H^{-1}}\,\F^{\mu\rho\,M}\F^{\nu}{}_{\rho\,M}+\frac{1}{8}\,e^{-2\Phi}\,\F_{\mu\nu}{}^{M}\left(\partial_{\rho}\H\partial^{\rho}\H^{-1}\right)_{M}{}^{N}\F^{\mu\nu}{}_{N}\\
      &\qquad+\frac{1}{4}\,e^{-2\Phi}\,\F_{\mu\nu}{}^{M}\left(\partial^{\mu}\H\partial_{\rho}\H^{-1}\right)_{M}{}^{N}\F^{\nu\rho}{}_{N}-e^{-2\Phi}\,\partial_{\mu}\Phi\,\F_{\nu\rho}{}^{M}\left(\partial^{\nu}\H\H^{-1}\right)_{M}{}^{N}\F^{\mu\rho}{}_{N}\\
      &\qquad+m\,e^{2\Phi}\,\epsilon^{\mu\nu\rho}\,\left(\Omega_{\mu\nu\rho}^{(\omega_{\rm E})}+\frac{1}{4}\,e^{-2\Phi}\,\F_{\mu\sigma}{}^{M}\left(\partial^{\sigma}\H_{MN}+2\,\partial^{\sigma}\Phi\,\H_{MN}\right)\F_{\nu\rho}{}^{N}\right)\bigg)\Bigg].
    \end{aligned}
  \end{equation}

  \subsection{Dualisation of the vector fields}
  We know dualise the vector fields into scalars by using the two-derivative dualisation eq.~\eqref{eq:dual2dermtheta} in the form
  \begin{equation}
    \F_{\mu\nu}{}^{M}=e^{2\Phi}\epsilon_{\mu\nu\rho}D^{\rho}\xi_{N}\H^{NM},
  \end{equation}
  with $D_{\mu}\xi_{M}=\partial_{\mu}\xi_{M}+m\,\A_{\mu\,M}$. We get
  \begin{align}
    \widetilde{I}''_{1} & = \int\d^{3}x\,\sqrt{-g_{\rm E}}\,e^{-2\Phi}\,\Bigg[\nonumber\\
    &-\frac{a+b}{8}\bigg(-\partial_{\mu}\Phi\partial^{\mu}\Phi\,\partial_{\nu}\Phi\partial^{\nu}\Phi+\frac{1}{16}\,\Tr{\partial_{\mu}\H\partial_{\nu}\H^{-1}\partial^{\mu}\H\partial^{\nu}\H^{-1}} \nonumber\\
    &\qquad+\frac{1}{32}\,\Tr{\partial_{\mu}\H\partial_{\nu}\H^{-1}}\Tr{\partial^{\mu}\H\partial^{\nu}\H^{-1}}-\frac{1}{64}\,\Tr{\partial_{\mu}\H\partial^{\mu}\H^{-1}}\Tr{\partial_{\nu}\H\partial^{\nu}\H^{-1}}\nonumber\\
    &\qquad+\frac{1}{4}\,e^{4\Phi}\,D_{\mu}\xi^{M}D_{\nu}\xi_{M}D^{\mu}\xi^{N}D^{\nu}\xi_{N}-\frac{1}{4}\,e^{4\Phi}\,D_{\mu}\xi_{M}\H^{MN}D_{\nu}\xi_{N}D^{\mu}\xi_{P}\H^{PQ}D^{\nu}\xi_{Q} \nonumber\\
    &\qquad+\frac{1}{4}\,e^{4\Phi}\,D_{\mu}\xi_{M}\H^{MN}D^{\mu}\xi_{N}D_{\nu}\xi_{P}\H^{PQ}D^{\nu}\xi_{Q}+\frac{1}{2}\,e^{2\Phi}\,D_{\mu}\xi_{M}\left(\H^{-1}\partial_{\nu}\H\partial^{\nu}\H^{-1}\right)^{MN}D^{\mu}\xi_{N}  \nonumber\\ 
    &\qquad-\frac{1}{2}\,e^{2\Phi}\,D_{\mu}\xi_{M}\left(\H^{-1}\partial^{\mu}\H\partial_{\nu}\H^{-1}\right)^{MN}D^{\nu}\xi_{N}-2\,e^{2\Phi}\,\partial_{\mu}\Phi\partial_{\nu}\Phi\,D^{\mu}\xi_{M}\H^{MN}D^{\nu}\xi_{N} \nonumber\\
    &\qquad+\frac{1}{2}\,\partial_{\mu}\Phi\partial_{\nu}\Phi\,\Tr{\partial^{\mu}\H\partial^{\nu}\H^{-1}}-\frac{1}{4}\,\partial_{\mu}\Phi\partial^{\mu}\Phi\,\Tr{\partial_{\nu}\H\partial^{\nu}\H^{-1}}  +e^{2\Phi}\,\partial_{\mu}\Phi \,D_{\nu}\xi_{M}\partial^{\mu}\H^{MN}D^{\nu}\xi_{N}\nonumber\\
    &\qquad+m\,e^{2\Phi}\,\epsilon^{\mu\nu\rho}\left(\frac{2}{3}\,\Omega_{\mu\nu\rho}^{\rm (GS)}-\frac{1}{2}\,e^{2\Phi}\,D_{\mu}\xi_{M}\left(\H^{-1}\partial_{\nu}\H\right)^{M}{}_{N}D_{\rho}\xi^{N}\right)-4\,m^{2}\,e^{4\Phi}\,\partial_{\mu}\Phi\partial^{\mu}\Phi-m^{4}\,e^{8\Phi}\bigg)\nonumber\\
    &+\frac{a-b}{4}\bigg(-\frac{1}{16}\,\Tr{\partial_{\mu}\H\partial^{\mu}\H^{-1}\partial_{\nu}\H\partial^{\nu}\H^{-1}\H\eta} + \frac{1}{4}\,e^{4\Phi}\,D_{\mu}\xi_{M}D_{\nu}\xi^{M}\,D^{\mu}\xi_{P}\H^{PQ}D^{\nu}\xi_{Q}\nonumber\\
    &\qquad-\frac{1}{4}\,e^{2\Phi}\,D_{\mu}\xi_{M}\left(\partial_{\nu}\H^{-1}\partial^{\mu}\H\right)^{M}{}_{N}D^{\nu}\xi^{N}-\frac{1}{2}\,e^{2\Phi}\,\partial_{\mu}\Phi\partial^{\mu}\Phi\,D_{\nu}\xi_{M}D^{\nu}\xi^{M} -\frac{1}{8}\,\Tr{\partial_{\mu}\H\partial_{\nu}\H^{-1}}\,D^{\mu}\xi_{M}D^{\nu}\xi^{M}\nonumber\\[-5pt]
    &\qquad+\frac{1}{16}\,\Tr{\partial_{\mu}\H\partial^{\mu}\H^{-1}}\,D_{\nu}\xi_{M}D^{\nu}\xi^{M}+e^{2\Phi}\,\partial_{\mu}\Phi\,D^{\mu}\xi_{M}\left(\partial_{\nu}\H^{-1}\H\right)^{M}{}_{N}D^{\nu}\xi^{N}\nonumber\\
    &\qquad+m\,e^{2\Phi}\,\epsilon^{\mu\nu\rho}\,\Omega_{\mu\nu\rho}^{(\omega_{\rm E})}\bigg)\Bigg], \label{eq:secondI1firstderiv}
  \end{align}
  which is equivalent to eq.~\eqref{eq:I1secondd+1currents}.
  Note that the terms $\frac{m}{4}\,\epsilon^{\mu\nu\rho}\F_{\mu\sigma}{}^{M}\left(\partial^{\sigma}\H_{MN}+2\,\partial^{\sigma}\Phi\,\H_{MN}\right)\F_{\nu\rho}{}^{N}$ give, upon dualisation,
  \begin{equation}
    -\frac{m}{2}\,e^{4\Phi}\,\epsilon^{\mu\nu\rho}D_{\mu}\xi_{M}\left(\partial_{\nu}\H^{MN}+2\,\partial_{\nu}\Phi\,\H^{MN}\right)D_{\rho}\xi_{N}=0.
  \end{equation}
  Idem for $m\,\epsilon^{\mu\nu\rho}\,e^{-2\Phi}\,\partial_{\sigma}\Phi\,\F_{\mu\nu}{}^{M}\F^{\sigma}{}_{\rho\,M}$.

\end{appendix}

\bibliography{refs}

\end{document}